\documentclass[aps,prl,showpacs,twocolumn,superscriptaddress,floatfix]{revtex4-2}
\usepackage[utf8]{inputenc}
\newcommand{\figurescale}{1}
\usepackage{verbatim}
\usepackage{amsmath}
\usepackage{mathtools}

\DeclarePairedDelimiterX\braket[2]{\langle}{\rangle}{#1 \delimsize\vert #2}

\usepackage[pdftex]{graphicx}
\usepackage[separate-uncertainty=true]{siunitx}
\DeclareSIUnit{\rpm}{rpm}
\usepackage[colorlinks=true,urlcolor=black,linkcolor=black,citecolor=black]{hyperref}
\usepackage[usenames,dvipsnames]{xcolor}
\usepackage[T1]{fontenc}
\usepackage{placeins}
\usepackage{nicefrac}
\usepackage{braket}
\usepackage{pdfcomment}
\usepackage{xcolor}
\usepackage{braket}
\usepackage[normalem]{ulem}

\newcommand{\bq}{\mathbf{q}}

\usepackage{lineno}

\begin{document}

\title{The bulk van der Waals layered magnet CrSBr is a quasi-1D material}
%
\author{Julian~Klein}\email{jpklein@mit.edu}
\affiliation{Department of Materials Science and Engineering, Massachusetts Institute of Technology, Cambridge, Massachusetts 02139, USA}
\author{Benjamin~Pingault}
\affiliation{John A. Paulson School of Engineering and Applied Sciences, Harvard University, Cambridge, Massachusetts 02138, USA}
\affiliation{QuTech, Delft University of Technology, 2600 GA Delft, The Netherlands}
\author{Matthias~Florian}
\affiliation{Department of Electrical and Computer Engineering,  Department of Physics, University of Michigan, Ann Arbor, Michigan 48109, United States}
\author{Marie-Christin~Heißenbüttel}
\affiliation{Institut für Festkörpertheorie, Westfälische Wilhelms-Universität Münster, 48149 Münster, Germany}
\author{Alexander~Steinhoff}
\affiliation{Institut für Theoretische Physik, Universität Bremen, P.O. Box 330 440, 28334 Bremen, Germany}
\affiliation{Bremen Center for Computational Materials Science, University of Bremen,
28359, Bremen, Germany}
\author{Zhigang~Song}
\affiliation{John A. Paulson School of Engineering and Applied Sciences, Harvard University, Cambridge, Massachusetts 02138, USA}
\author{Kierstin~Torres}
\affiliation{Department of Materials Science and Engineering, Massachusetts Institute of Technology, Cambridge, Massachusetts 02139, USA}
\author{Florian~Dirnberger}
\affiliation{Department of Physics, City College of New York, New York, NY 10031, USA}
\author{Jonathan~B.~Curtis}
\affiliation{College of Letters and Science, UCLA, Los Angeles, CA 90095 USA}
\affiliation{John A. Paulson School of Engineering and Applied Sciences, Harvard University, Cambridge, Massachusetts 02138, USA}
\author{Mads~Weile}
\affiliation{Center for Visualizing Catalytic Processes (VISION), Department of Physics, Technical University of Denmark, DK-2800 Kgs. Lyngby, Denmark}
\author{Aubrey~Penn}
\affiliation{MIT.nano, Massachusetts Institute of Technology, Cambridge, Massachusetts 02139, USA}
\author{Thorsten~Deilmann}
\affiliation{Institut für Festkörpertheorie, Westfälische Wilhelms-Universität Münster, 48149 Münster, Germany}
\author{Rami~Dana}
\affiliation{Department of Materials Science and Engineering, Massachusetts Institute of Technology, Cambridge, Massachusetts 02139, USA}
\author{Rezlind~Bushati}
\affiliation{Department of Physics, City College of New York, New York, NY 10031, USA}
\affiliation{Department of Physics, The Graduate Center, City University of New York, New York, NY 10016, USA}
\author{Jiamin~Quan}
\affiliation{Photonics Initiative, CUNY Advanced Science Research Center, New York, NY, 10031, USA}
\affiliation{Physics Program, Graduate Center, City University of New York, New York, NY, 10026, USA}
\author{Jan~Luxa}
\affiliation{Department of Inorganic Chemistry, University of Chemistry and Technology Prague, Technická 5, 166 28 Prague 6, Czech Republic}
\author{Zdenek~Sofer}
\affiliation{Department of Inorganic Chemistry, University of Chemistry and Technology Prague, Technická 5, 166 28 Prague 6, Czech Republic}
\author{Andrea~Al\`{u}}
\affiliation{Photonics Initiative, CUNY Advanced Science Research Center, New York, NY, 10031, USA}
\affiliation{Physics Program, Graduate Center, City University of New York, New York, NY, 10026, USA}
\author{Vinod~M.~Menon}
\affiliation{Department of Physics, City College of New York, New York, NY 10031, USA}
\affiliation{Department of Physics, The Graduate Center, City University of New York, New York, NY 10016, USA}
\author{Ursula~Wurstbauer}
\affiliation{Institute of Physics and Center for Nanotechnology, University of Münster, 48149 Münster, Germany}
\author{Michael~Rohlfing}
\affiliation{Institut für Festkörpertheorie, Westfälische Wilhelms-Universität Münster, 48149 Münster, Germany}
\author{Prineha~Narang}\email{prineha@ucla.edu}
\affiliation{College of Letters and Science, UCLA, Los Angeles, CA 90095 USA}
\affiliation{John A. Paulson School of Engineering and Applied Sciences, Harvard University, Cambridge, Massachusetts 02138, USA}
\author{Marko~Lon\v{c}ar}\email{loncar@seas.harvard.edu}
\affiliation{John A. Paulson School of Engineering and Applied Sciences, Harvard University, Cambridge, Massachusetts 02138, USA}
\author{Frances~M.~Ross}\email{fmross@mit.edu}
\affiliation{Department of Materials Science and Engineering, Massachusetts Institute of Technology, Cambridge, Massachusetts 02139, USA}
%
%
%
\date{\today}
%
%
\begin{abstract}
\textbf{Abstract.}
Correlated quantum phenomena in one-dimensional (1D) systems that exhibit competing electronic and magnetic order are of strong interest for studying fundamental interactions and excitations, such as Tomonaga-Luttinger liquids and topological orders and defects with properties completely different from the quasiparticles expected in their higher-dimensional counterparts. However, clean 1D electronic systems are difficult to realize experimentally, particularly magnetically ordered systems. Here, we show that the van der Waals layered magnetic semiconductor CrSBr behaves like a quasi-1D material embedded in a magnetically ordered environment. The strong 1D electronic character originates from the Cr-S chains and the combination of weak interlayer hybridization and anisotropy in effective mass and dielectric screening with an effective electron mass ratio of $m^e_X/m^e_Y \sim 50$. This extreme anisotropy experimentally manifests in strong electron-phonon and exciton-phonon interactions, a Peierls-like structural instability and a Fano resonance from a van Hove singularity of similar strength of metallic carbon nanotubes. Moreover, due to the reduced dimensionality and interlayer coupling, CrSBr hosts spectrally narrow ($\SI{1}{\milli\electronvolt}$) excitons of high binding energy and oscillator strength that inherit the 1D character. Overall, CrSBr is best understood as a stack of weakly hybridized monolayers and appears to be an experimentally attractive candidate for the study of exotic exciton and 1D correlated many-body physics in the presence of magnetic order.
\end{abstract}
%
%
\maketitle
%
%


Experimental realizations of one-dimensional (1D) platforms for studying quantum phenomena are rare, some examples being ultra-cold quantum gases,~\cite{Paredes.2004} atomic chains~\cite{Blumenstein.2011} or carbon nanotubes.~\cite{Rao.1997,Saito.1998,Bockrath.1999,Brown.2001,Nguyen.2007,Farhat.2007} While such platforms are essential for the study of correlated phenomena like Tomonaga-Luttinger liquids,~\cite{Bockrath.1999,Blumenstein.2011} 1D spin chains~\cite{Lake.2005} or for realizing quantum conducting wires,~\cite{Langer.1996} they may be unstable, hard to scale up or suffer from ensemble inhomogeneities.

The recent emergence of the field of quasi-1D materials represents a technologically appealing alternative.~\cite{Balandin.2022} Quasi-1D materials are ideal systems to realize, study and exploit strongly correlated phenomena in a solid-state platform. They are typically characterized by an anisotropy of the structural atomic arrangement and the related electronic intra-planar interactions in different crystallographic directions. Examples include MoS$_2$ nanotubes~\cite{Remskar.2001} and hBN nanotubes,~\cite{Golberg.2007} III-V-based nanowires~\cite{Pfeiffer.1993,Meier.2021} or other quasi-1D materials like ZrTe$_3$,~\cite{Felser.1998} NbSe$_3$~\cite{Meerschaut.1986} or TiSe$_3$.~\cite{Jin.2015} Low-dimensional systems with both electronic and magnetic character are particularly intriguing and several efforts have been dedicated to their design and fabrication.~\cite{Deshpande.2010} Early examples include dilute magnetic semiconductors like Mn-doped quantum wells and quantum wires that exhibit electronic and optical excitations that are strongly correlated with the magnetic degree of freedom.~\cite{Furdyna.1988,Wurstbauer.2010}

Quasi-1D systems with both magnetic and electronic properties are expected to offer an even greater range of exotic quantum phenomena and functionalities. The recent observation of long-range magnetic order in van der Waals magnets~\cite{Gong.2017,Huang.2017} provides a prospect for experimental realization of materials with correlated excitations including electron-electron, electron-spin, exchange and electron-phonon interactions. In one such magnet, TiOCl, Spin-Peierls and CDW physics has been reported.~\cite{Shaz.2005,BlancoCanosa.2009,Prodi.2010} TiOCl is a member of the group of chalcogen-halides that are described by the stoichiometric formula MXP composed of a transition metal (M = Cr, Fe, V), a chalcogen (X = S, Se, O) and a halogen (P = Br, Cl, I). Without including spin, the chalcogen-halides have space group 59 (Pmmn) and point group $D_{2h}$. The observation of correlated physics in TiOCl suggests that electronic 1D character may be visible in materials that have similar structure but even larger structural anisotropy.

\begin{figure*}
	\scalebox{\figurescale}{\includegraphics[width=1\linewidth]{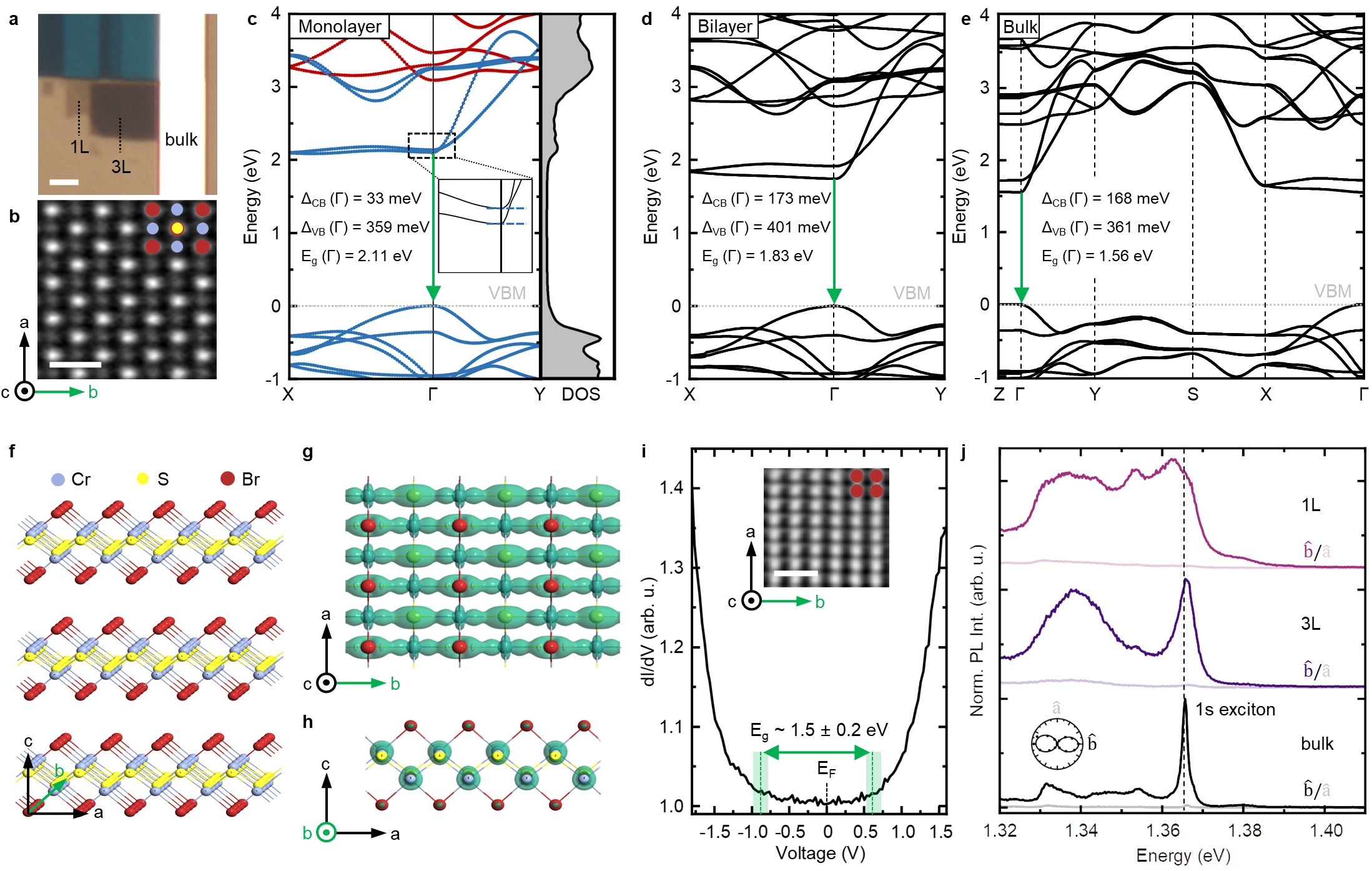}}
	\renewcommand{\figurename}{FIG.|}
	\caption{\label{fig1}
		\textbf{Quasi-1D electronic structure of CrSBr in the bulk.}
		\textbf{a}, Optical micrograph of mechanically exfoliated CrSBr. Needle-like crystals extend along the $a$-direction. Scale bar is $\SI{4}{\micro\meter}$.
		\textbf{b}, STEM-HAADF image of multilayer CrSBr taken at an electron beam energy of $\SI{200}{\kilo\electronvolt}$. Scale bar is $\SI{0.5}{\nano\meter}$.
		\textbf{c}, 1L CrSBr DFT-$GW$ calculation of the high symmetry points $X- \Gamma - Y$ showing a flat conduction band along the $\Gamma - X$ direction (effective electron mass $m_X^e = 7.31 m_0$) corresponding to the $a$-direction and a highly dispersive band along the $\Gamma-Y$ direction ($m_Y^e = 0.14 m_0$) corresponding to the $b$-direction. The blue and red colors represent the majority and minority (in-plane) spin polarization. The corresponding density of states (DOS) is shown. Inset: Zoom in of the split conduction bands at $\Gamma$ ($\Delta_\text{CB} = 33$\,meV).
		\textbf{d}-\textbf{e}, DFT-$GW$ calculations of the 2L and bulk CrSBr.
		\textbf{f-h}, Schematic illustration of bulk CrSBr with the relevant electronic structure that is formed by mainly Cr and S atomic orbitals. The charge density of states shown near the Fermi level is obtained by DFT calculations. The charge density includes the states in the range $\SI{0.1}{\electronvolt}$ below the top of the valence band and $\SI{0.1}{\electronvolt}$ above the bottom of the conduction band. Although the atomic distance in the $a$-direction is less than in the $b$-direction, the electronic states form chains (charge density in green) along the $b$-direction that are only weakly hybridized (coupled) along the $a$-direction.
		\textbf{i}, Room temperature constant height dI/dV from STS ($I_t = \SI{50}{\pico\ampere}$ at $V_{bias}=\SI{+0.2}{\volt}$) with a single-particle gap of $\sim 1.5\pm 0.2 \SI{}{\electronvolt}$. Green error bars show the uncertainty from the CBM and VBM onset. Inset: Topographic STM image of CrSBr surface showing the Br atoms on the surface. Scale bar is $\SI{2}{\nano\meter}$.
		\textbf{j}, Low-temperature ($\SI{4.2}{\kelvin}$) PL of 1L, 3L and bulk ($\SI{36.8}{\nano\meter} \sim$ 46L) CrSBr for the electric field co-polarized in excitation and detection along the $b$-and $a$-direction, respectively. Excitation power is $\SI{100}{\micro\watt}$. The PL emission of the 1s exciton in bulk CrSBr is highlighted. Inset: Anisotropic 1s exciton emission of bulk CrSBr.
		}
\end{figure*}

\begin{figure*}
	\scalebox{\figurescale}{\includegraphics[width=1\linewidth]{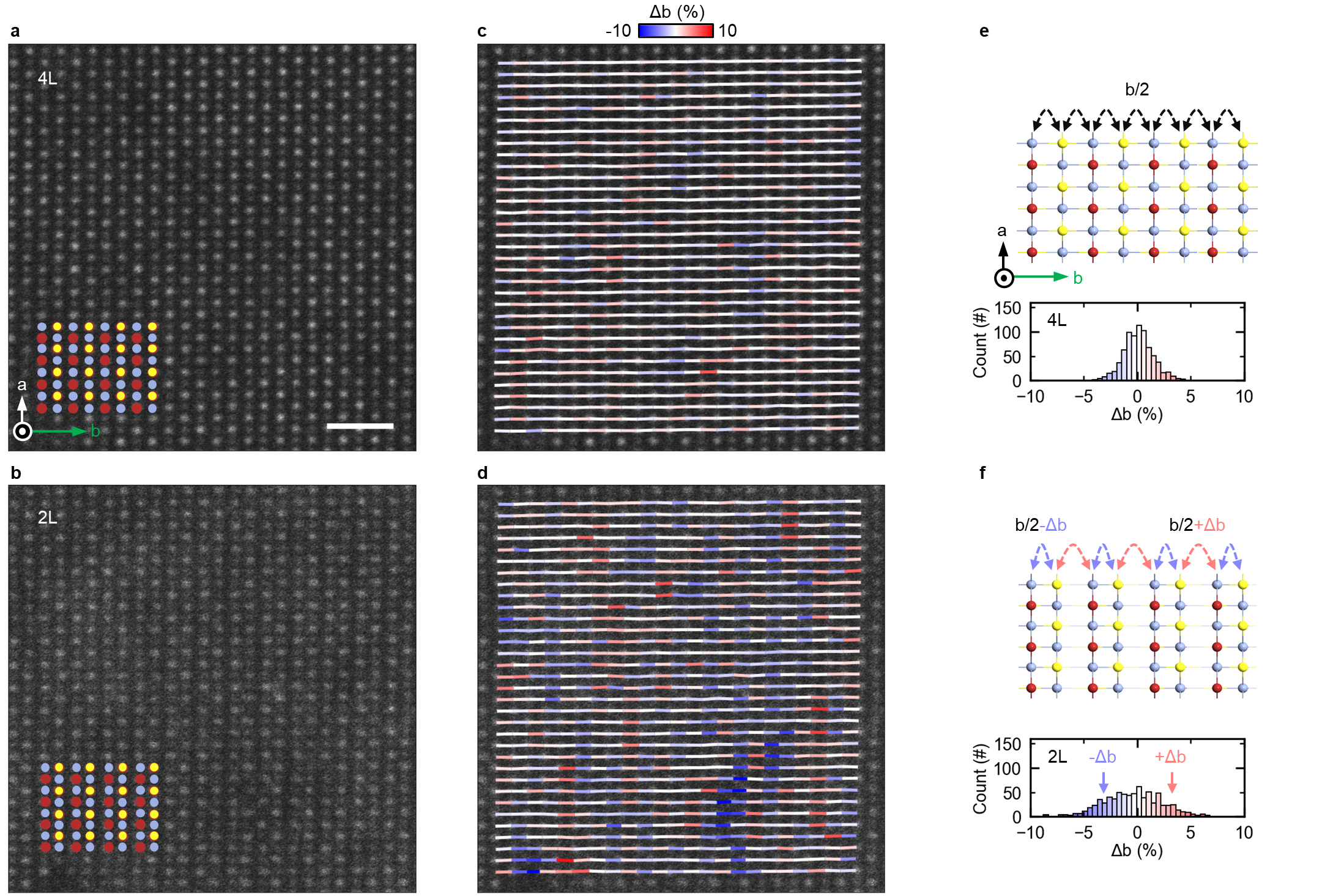}}
	\renewcommand{\figurename}{FIG.|}
	\caption{\label{fig2}
		\textbf{Anisotropic structural distortion along the $b$-direction in CrSBr.}
	\textbf{a,b}, STEM-HAADF image of a 4L and 2L CrSBr flake. Image was taken at room temperature ($\SI{300}{\kelvin}$), a beam energy of $\SI{200}{\kilo\electronvolt}$ and a beam current of $\SI{40}{\pico\ampere}$. The inset shows the atomic arrangement. Scale bar is $\SI{1}{\nano\meter}$.
        \textbf{c,d}, Corresponding visualization of the percentage change of inter-atomic distance $\Delta b$ between the Cr and S/Br atomic columns along the $b$-direction. The distances are normalized to the pristine nondistorted distance $b/2$ of a 6L CrSBr.
        \textbf{e}, Schematic illustration of the nondistorted atomic arrangement without a structural distortion as further corroborated from the histogram of 4L CrSBr obtained from \textbf{c}.
        \textbf{f}, Schematic illustration of the atomic distortion along the $b$-direction. The absolute lattice distances along the $b$-direction are characterized by $b/2 \pm \Delta b$. The corresponding histogram of a 2L CrSBr from the analysis in \textbf{d} shows additional characteristic shoulders at $\pm \Delta b \sim \pm 4\%$ from the alternating twin lines.
		}
\end{figure*}

One such material is CrSBr. ~\cite{Katscher.1966,Gser.1990,Wang.2019,Wang.2020,Telford.2020,Lee.2021,Wilson.2021,Yang.2021,Klein.2021,Klein.2022,Telford.2022,Wu.2022,Torres.2023} This is a semiconductor with an energy gap of $\sim \SI{1.5}{\electronvolt}$.~\cite{Telford.2020,Wang.2019} It is also an A-type antiferromagnet (AFM) with the magnetic easy-axis along the $b$-direction and a high N\'{e}el temperature of $T_N \sim \SI{132}{\kelvin}$.~\cite{Gser.1990,Telford.2020} Significantly, CrSBr is a relatively stable material in air.~\cite{Torres.2023} Recent works have shown structural phase transformations,~\cite{Klein.2021} correlated magneto-optical~\cite{Wilson.2021,Klein.2022,Torres.2023} and magneto-transport~\cite{Telford.2020,Telford.2022,Wu.2022} properties and the coupling of excitons and magnons.~\cite{Bae.2022} Moreover, a recent transport study has reported an unusual anisotropy in the electron conductivity along the $a$- and $b$-directions of CrSBr with striking ratios of up to $10^5$, attributed to quasi-1D transport.~\cite{Wu.2022}

Based on its stability and the optical, electronic and magnetic properties, we believe that CrSBr is an outstanding candidate for practical engineering of atomic-scale magnetic excitations that offer opportunities for nano-spintronic, nano-spin-photonic and memory devices. However, it is essential to first understand the origin of the peculiar electronic structure of CrSBr and its impact on the material's vibrational, optical and magnetic properties and related quasiparticle interactions. In this manuscript, we have therefore examined the band structure, atomic structure, quasiparticles and their mutual couplings in CrSBr both experimentally and theoretically. Our key finding is that CrSBr is unambiguously a quasi-1d material even in bulk exfoliated form. We explain this result by suggesting that CrSBr exhibits a strong 1D electronic character which is due to electronic chains that are weakly coupled within each layer in combination with a weak interlayer hybridization and anisotropy in effective mass and dielectric screening. This is further corroborated by our observation of a structural distortion along the direction of high electronic density which is suggestive of a Peierls instability. Moreover, we find that all quasiparticle excitations and their mutual couplings that we measure experimentally, using resonant and non-resonant Raman, low-temperature photoluminescence, magneto-reflectivity and photoluminescence excitation spectroscopy, can be understood through a dominant role of this low-dimensional electronic structure. Strikingly, optical spectroscopy on exfoliated bulk CrSBr shows ultra-clean excitonic signatures with Lorentzian linewidths of only $\SI{1}{\milli\electronvolt}$, usually known only from high quality boron nitride encapsulated monolayer semiconductors like MoS$_2$ or WSe$_2$; we attribute this also to the 1D electronic character, in combination with the absence of sample inhomogeneity from strain or substrate/surface effects that dominate measurements from other materials that can only be measured as atomically thin layers. From our results, we propose that bulk CrSBr is best understood as a stack of weakly coupled monolayers that host quasi-1D excitons of high binding energy with strong quasiparticle interactions due to the quasi-1D electronic character in each layer. Our work establishes bulk CrSBr as an ideal platform to explore rich quasiparticle excitations and their mutual couplings in a magnetically ordered environment.

\section{Results and Discussion}

\textbf{Quasi-1D electronic structure of CrSBr.} Unlike the previously investigated archetypal magnets CrI$_3$ and Cr$_2$Ge$_2$Te$_6$,~\cite{Huang.2017,Gong.2017} CrSBr shows good air-stability and can be easily cleaved down to a monolayer (1L) (see Fig.~\ref{fig1}a).~\cite{Torres.2023} Exfoliated crystals have a needle-like shape and exhibit structural anisotropy apparent from the atomic arrangement in a scanning transmission electron microscopy high-angle annular dark-field (STEM-HAADF) image (see Fig.~\ref{fig1}b).~\cite{Klein.2021} The strong electronic anisotropy is directly apparent from our 1L, 2L and bulk density functional theory (DFT-$GW$) calculation (see Fig.~\ref{fig1}c-e). The 1L calculation shows two conduction bands (energy splitting of $\SI{33}{\milli\electronvolt}$ at the $\Gamma$ point) that are flat along the $a$-direction but highly dispersive along the $b$-direction. Here the $a$- and $b$-directions are defined as $\Gamma - X$ and $\Gamma - Y$, respectively. The relevant electronic structure around the $\Gamma$-point can also be well described in a simple three-band $k\cdot p$ model (see SI). The $\Gamma$-point effective masses of the lowest conduction band and the topmost valence band in $\Gamma-X$- and $\Gamma-Y$-directions, respectively, are extracted from the DFT+$GW$ band structures: $m^{\textrm{e}}_{X}= 7.31 m_0$, $m^{\textrm{e}}_{Y}= 0.14 m_0$, $m^{\textrm{h}}_{X}= 2.84 m_0$, $m^{\textrm{h}}_{Y}= 0.45 m_0$. This manifests as a pronounced effective electron mass anisotropy of $\frac{m^e_X}{m^e_Y} \sim 50$. This mass anisotropy is significantly higher as compared to anisotropic black phosphorous.~\cite{Tran.2014} The electronic anisotropy originates from the orbital composition of the electronic bands around the conduction band minimum (CBM) and the valence band maximum (VBM). This is best visualized by the charge density around the Fermi level of CrSBr with predominant admixture from the Cr-S chains along the $b$-direction (see Fig.~\ref{fig1}g and h). The orbitals of the Br atoms only weakly admix into the conduction bands, resulting in the weakened interlayer hybridization. In the case of the 1L, we obtain admixtures to the lower conduction band of 64\% Cr, 32\% S and 4\% Br and for the upper conduction band 86\% Cr, 9\% S and 5\% Br. It is the dominant admixture of the Cr and S orbitals in the band structure that cause CrSBr's strong 1D electronic character, with a shallow 1D quantum confinement along the $a$-direction due to a weak intra-layer hybridization (coupling) of the chains. This weak hybridization is also apparent from the DFT-$GW$ calculation of bilayer (2L) and bulk CrSBr (see Fig.~\ref{fig1}d and e) and differs from semiconducting TMDCs.~\cite{Mak.2010,Splendiani.2010} We obtain a qualitatively similar band structure for 1L, 2L and bulk along the $X-\Gamma -Y$ direction, as expected from the weak interlayer hybridization. The fundamental band gap is direct and situated at the $\Gamma$ point, in addition to a nearly degenerate conduction band state at the $X$-point for the 1L. In general, we find that the effect of the flat band along $\Gamma - X$ becomes even more pronounced in the $GW$ calculation. Moreover, the strong mass anisotropy results in a large density of states (DOS) along the $\Gamma-X$ direction at the band edge which is expected to be accompanied by a van Hove singularity (see SI). However, it should be noted that this is not directly apparent from the calculated DOS (see Fig.~\ref{fig1}c) owing to the admixture of bands from all momenta making this effect less pronounced.

\begin{figure*}
	\scalebox{\figurescale}{\includegraphics[width=1\linewidth]{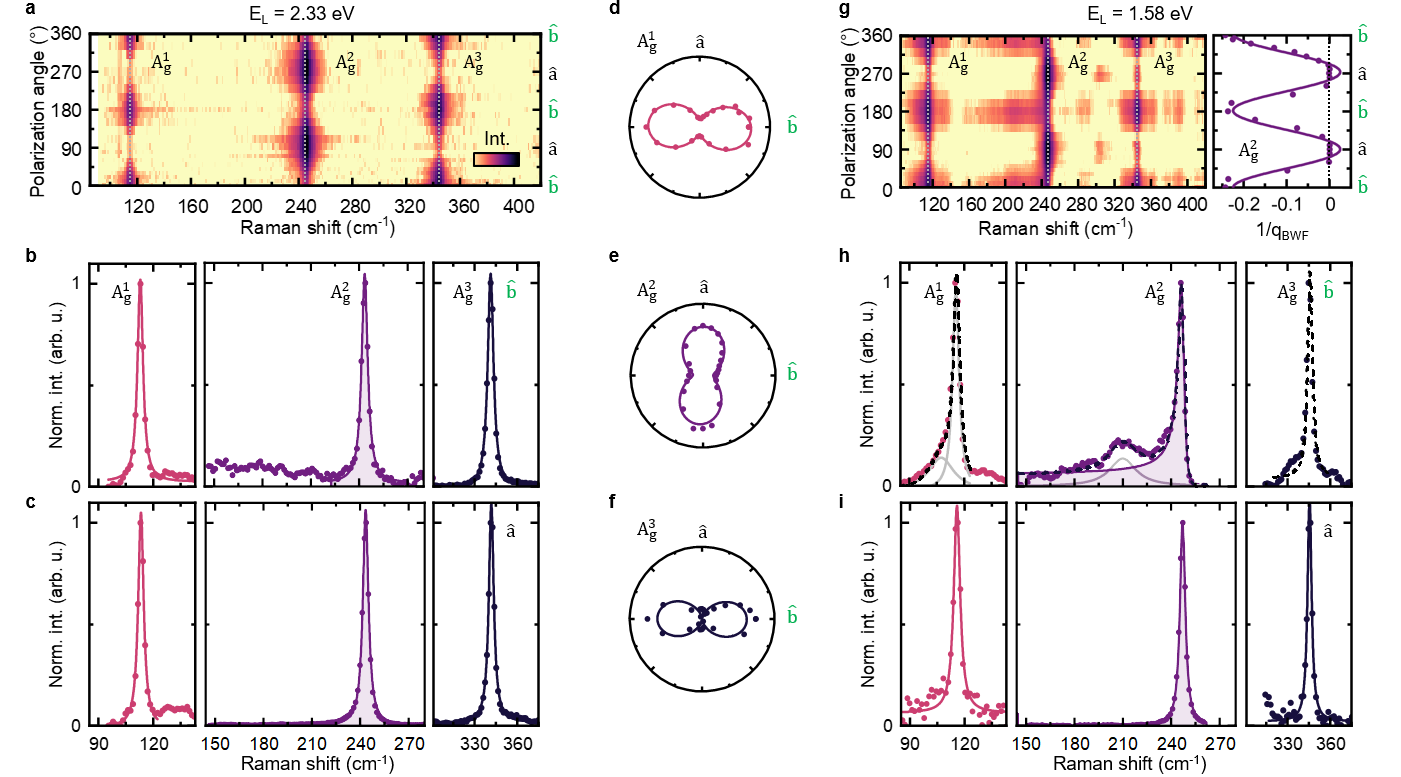}}
	\renewcommand{\figurename}{FIG.|}
	\caption{\label{fig3}
		\textbf{Pronounced and quasi-1D electron-phonon coupling in bulk CrSBr.}
		\textbf{a}, False color mapping of the polarization angle dependent Raman spectra with off-resonant excitation at $E_L = \SI{2.33}{\electronvolt}$ far above the single-particle gap for an excitation power of $\SI{250}{\micro\watt}$ at room temperature. The three main phonon modes $A^1_g$ ($\SI{115}{\per\centi\meter}$), $A^2_g$ ($\SI{245.5}{\per\centi\meter}$) and $A^3_g$ ($\SI{344}{\per\centi\meter}$) in CrSBr are highlighted. 
		\textbf{b}, Normalized Raman spectra for each individual mode with polarization along the $b$-direction (\^{b}, $\SI{0}{}^{\circ}$ or $\SI{180}{}^{\circ}$) shows Lorentzian line shapes as does \textbf{c} the $a$-direction (\^{a}, $\SI{90}{}^{\circ}$ or $\SI{270}{}^{\circ}$).
		\textbf{d-f}, Polarization angle dependent intensity of the $A^1_g$, $A^2_g$ and $A^3_g$ phonon mode for excitation at $E_L = \SI{1.58}{\electronvolt}$ and an excitation power of $\SI{1000}{\micro\watt}$. Only mode $A^2_g$ exhibits an intensity maximum along the \^{a}-direction. The solid lines are $\sin^2$ fits.
		\textbf{g}, Left: Polarization dependent Raman at an excitation energy of $E_L = \SI{1.58}{\electronvolt}$ that is in electronic resonance with the band gap. Right: The Breit-Wigner-Fano coupling parameter $1/q_{BWF}$ of the $A^2_g$ phonon mode oscillates between the $b$- and $a$-direction as further indicated by the sine fit.
		\textbf{h}, Normalized Raman spectra for each individual mode with the electric field co-polarized along the $b$-direction. The $A^2_g$ mode is fitted with a BWF resonance (see Eq.~\ref{eq:BWF}).
		\textbf{i}, Raman spectrum along the $a$-direction. All modes exhibit fully Lorentzian lineshapes.
		}
\end{figure*}

In order to access the electronic states experimentally, we perform scanning tunneling microscopy (STM) and spectroscopy (STS) at room temperature ($T = \SI{300}{\kelvin}$) on bulk CrSBr crystals cleaved under ultra high vacuum to obtain a clean surface. Figure~\ref{fig1}i shows a typical dI/dV trace. The extracted size of the band gap, $E_g = 1.5 \pm 0.2 \SI{}{\electronvolt}$, is consistent with the bulk DFT-$GW$ calculation yielding an energy gap of $\SI{1.56}{\electronvolt}$ (see Fig.~\ref{fig1}e) and in agreement with a previous STS study.~\cite{Telford.2020} However, the position of the Fermi level $E_F$ in our material suggests a more intrinsic behavior likely from a lower density of defects.~\cite{Telford.2020}

Low-temperature ($\SI{4.2}{\kelvin}$) PL (see Fig.~\ref{fig1}j) confirms the weak interlayer hybridization of CrSBr, in line with the other findings. The optical emission of CrSBr is strongly linearly polarized along the $b$-direction reflecting the electronic anisotropy. The spectra from 1L, 3L and bulk ($\SI{36.8}{\nano\meter} \sim$ 46L) show a clear signature of the 1s exciton recombination at $\sim\SI{1.366}{\electronvolt}$ along with other hitherto unidentified peaks below the 1s exciton in energy. Most importantly, the features in all spectra remain the same upon layer number reduction, but display significant line broadening that likely originates from disorder-related effects such as strain, unintentional doping from the substrate, defects or adsorbates. For bulk CrSBr, we obtain Lorentzian linewidths of the 1s exciton of $\sim \SI{1}{\milli\electronvolt}$, reminiscent of high-quality hBN encapsulated TMDCs.~\cite{Cadiz.2017,Shree.2018}\\

\textbf{Anisotropic structural distortion.} Next, we collect STEM-HAADF images of exfoliated CrSBr flakes from the bulk thickness down to the monolayer (see 2L and 4L in Fig.~\ref{fig2} and additional layer numbers in the SI) in order to probe structural changes that arise from the anisotropic electronic structure of CrSBr. Strikingly, we observe a structural dimerization of the crystal lattice along the $b$-direction forming twin lines as the layer number is reduced (see Fig.~\ref{fig2}c). This is not just a simple surface effect since we do not observe this structural distortion in STM topographic images (see inset Fig.~\ref{fig1}i).~\cite{Klein.2022} Importantly, this coincides with the direction of the strong quasi-1D electronic structure of CrSBr (see Fig.~\ref{fig1}g). We further visualize and quantify this structural distortion by analyzing the percentage change in the inter-atomic lattice distance $\Delta b$ (see Methods) between neighboring Cr and S/Br atomic columns along the $b$-direction (see Fig.~\ref{fig2}c and d). The corresponding extracted percentage change in the inter-atomic distances along the $b$-direction is shown in histograms in Fig.~\ref{fig2}e and f. We observe a significant structural distortion that becomes increasingly more dominant when the layer number of CrSBr is reduced. The bilayer exhibits a broadened distribution with characteristic shoulders at $\pm \Delta b$ that reflect the expansion and contraction of the chains. We obtain large structural distortions of $\Delta b \pm 4\%$, comparable to highly correlated materials like 1T-TaS$_2$.~\cite{Spijkerman.1997} All histograms remain centered around $0\%$ for all layer numbers suggesting no build up of net strain. Notably, for the 1L, we observe almost fully isolated twin lines that form 1D-like nanowire structures (see SI). Moreover, there is no structural distortion along the $a$-direction (see SI). 

The formation of twin lines along the $b$-direction is a hallmark signature of a quasi-1D material suggestive of a Peierls distortion.~\cite{Balandin.2022} This observation further supports the strong electronic anisotropy of CrSBr visible from a change in the atomic arrangement. The magnitude of the structural distortion is directly related to the strength of the electron-phonon coupling of the system.~\cite{Johannes.2008} The high value ($\pm 4\%$) observed for CrSBr is suggestive of very large electron-phonon coupling effects. We note, that the exact amount of lattice distortion is likely to further depend on pressure and temperature. The intricate interdependence of layer thickness, strain and electronic structure further provides fertile grounds for future studies.

\textbf{Quasi-1D electron-phonon coupling.} The pronounced quasi-1D electronic structure and band anisotropy is expected to manifest itself in the quasiparticle excitations and their mutual interactions. From the observed structural distortion suggestive of a Peierls instability, we already expect a strong electron-phonon coupling in CrSBr. To explore such interactions between excitations of the 2D lattice and excitations of the quasi-1D electronic system, we perform resonant inelastic light scattering (RILS) spectroscopy.~\cite{Loudon.1978} In order to selectively probe the quasiparticles of the lattice excitations, we use non-resonant Raman with an excitation well above the band gap of $E_L =\SI{2.33}{\electronvolt}$; to study the interaction between the quasiparticles' electronic inter-band, intra-band and lattice excitations, e.g. electron-phonon and exciton-phonon, we perform resonant Raman with excitation close to the single-particle band edge of CrSBr ($E_L = \SI{1.58}{\electronvolt}$). For this measurement, we co-align linear excitation and detection polarizations and collect spectra of a bulk CrSBr flake ($\sim \SI{20}{\nano\meter}$) as a function of orientation with respect to the crystal axes.

The contour map for non-resonant excitation ($E_L = \SI{2.33}{\electronvolt}$) shows three main modes that we identify as the $A^1_g$ ($\SI{115}{\per\centi\meter}$), $A^2_g$ ($\SI{245.5}{\per\centi\meter}$) and $A^3_g$ ($\SI{344}{\per\centi\meter}$) modes (see Fig.~\ref{fig3}a) analogously to the modes observed in CrOCl~\cite{Zhang.2019} While all modes are due to out-of-plane atomic displacements, the $A^1_g$ and $A^3_g$ modes exhibit an intensity maximum along the $b$-direction while the $A^2_g$ mode exhibits a maximum along the $a$-direction (see Fig.~\ref{fig3}d-f). Importantly, the lineshape of all modes is Lorentzian for all polarization angles.

For resonant excitation ($E_L = \SI{1.58}{\electronvolt}$), we observe a number of additional resonant modes with distinct energies and a very pronounced asymmetric line shape of the $A^2_g$ mode (see Fig.~\ref{fig3}g-i) but only when the electric field vector is aligned along the $b$-direction (k-vector pointing along the $a$-direction). This asymmetry is absent for the same experimental conditions in CrOCl ruling out a laser induced origin (see SI). This line shape can be interpreted as a signature of pronounced electron-phonon interactions in presence of resonance effects, e.g. coupling between lattice and electronic excitations.~\cite{Rao.1997,Saito.1998} A potential interpretation is a strong polarization-selective 1D-like electron-phonon interaction of the $A^2_g$ mode with the electronic structure. The $A^2_g$ mode is polarized along the $a$-direction and is therefore prone to strongly couple to the high DOS of the electronic system along the $\Gamma - X$ direction (see Fig.~\ref{fig1}c). This is known as a Breit-Wigner-Fano (BWF) resonance, with the phonon as discrete state and the 1D DOS along the $a$-direction as the electronic continuum from shallow 1D quantum confinement.~\cite{Rao.1997,Saito.1998} The spectral form of a BWF resonance is given by

\begin{equation}
\label{eq:BWF}
    I(\omega) = I_0 \frac{[1 + (\omega - \omega_{BWF}/q_{BWF}\Gamma)]^2}{1 + [(\omega - \omega_{BWF})/\Gamma]^2}
\end{equation}

where the asymmetry factor $1/q_{BWF}$ characterizes the coupling strength between the phonon and the electronic continuum and $\omega_{BWF}$ is the uncoupled BWF peak frequency. The BWF resonance profile reproduces the experimental $A^2_g$ mode spectrum under resonant excitation along the $b$-direction (see Fig.~\ref{fig3}b). The BWF coupling parameters $1/q_{BWF}$ deduced from those fits (see right panel, Fig.~\ref{fig3}g) reveal a high coupling strength between phonons and electronic continuum of $1/q_{BWF} = -0.24$ for polarization along the $b$-direction. This is comparable with reports on metallic CNTs with values between $-0.2$ and $-0.6$.~\cite{Brown.2001,Nguyen.2007,Farhat.2007} In contrast, the line shape of the RILS spectra of $A^2_g$ measured along the $a$-direction is best described by a pure Lorentzian with $1/q_{BWF}$ yielding vanishing coupling between phonons and the electronic continuum. The observation of the highly anisotropic Fano resonance can serve as a strong signature for the 1D nature of the continuum.~\cite{Brown.2001,Nguyen.2007,Farhat.2007} While the observation of a Fano lineshape is surprising in a semiconductor like CrSBr, such an observation is not limited to metallic systems like CNTs but also observed in gapped materials that generally exhibit pronounced electron-phonon interactions.~\cite{Zhang.2011,Tan.2017} CrSBr, has shown a high intrinsic carrier concentration of $\sim 10^{13}\SI{}{\per\centi\meter\squared}$~\cite{Telford.2022} in agreement with a high concentration of Br vacancy defects~\cite{Klein.2022} that induce n-doping in the material shifting the Fermi level close to the conduction band. Moreover, low-dimensional behavior and Fano-physics~\cite{Caimi.2004} has also been observed in TiOCl suggesting that this is likely a universal effect in materials that are of FeOCl-type. However, other possible interpretations exist. For example, the lineshape can be affected by resonance modes from different points in momentum space. Additional insight into the Raman process is required to further elucidate the physics of the lineshape of the $A^2_g$ mode and the complex resonance Raman spectrum of CrSBr.

\begin{figure*}
	\scalebox{\figurescale}{\includegraphics[width=1\linewidth]{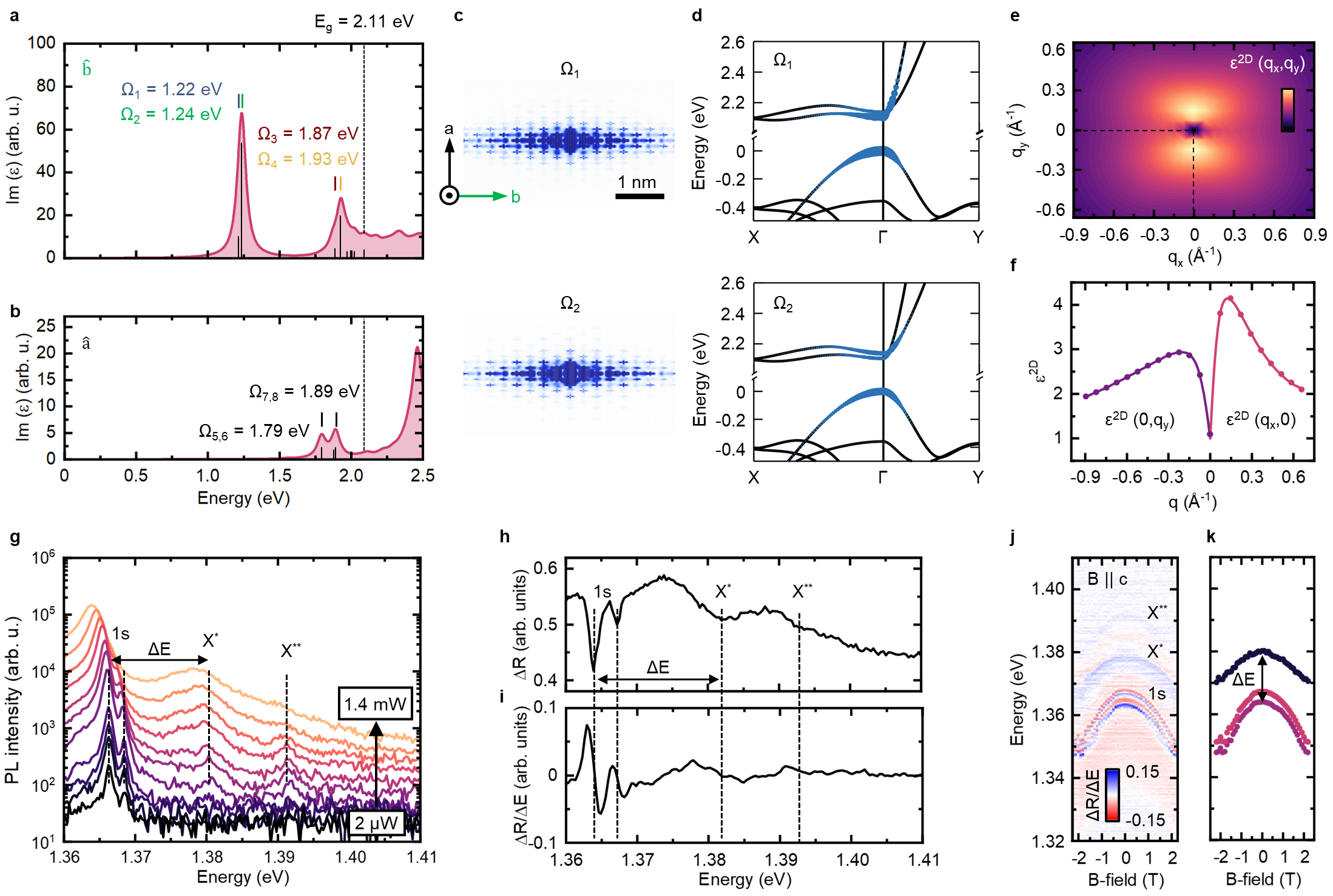}}
	\renewcommand{\figurename}{FIG.|}
	\caption{\label{fig4}
	    \textbf{Quasi-1D excitons with high binding energy from dielectric and mass anisotropy in CrSBr.}
	    \textbf{a}, Calculated optical absorption spectrum obtained ab initio by solving the BSE of 1L CrSBr with an electric field polarized along the $b$-direction ($\Gamma -Y$-direction). The spectrum shows four excitonic resonances. $\Omega_1$ and $\Omega_2$ originate from the two split conduction bands with a small energy splitting of $\Delta E_{\Omega_1,\Omega_2} \sim \SI{20}{\milli\electronvolt}$. The binding energies are given by $\SI{0.88}{\electronvolt}$ and $\SI{0.90}{\electronvolt}$ for $\Omega_1$ and $\Omega_2$, respectively. $\Omega_3$ and $\Omega_4$ are excitons from energetically higher lying bands.
	    \textbf{b}, Calculated optical absorption spectrum from ab initio BSE of 1L CrSBr for the electric field polarized along the $a$-direction ($\Gamma -X$-direction) showing the strong electronic anisotropy and the absence of $\Omega_1$ and $\Omega_2$.
	    \textbf{c}, Corresponding anisotropic real space exciton wavefunction of $\Omega_1$ and $\Omega_2$. 
	    \textbf{d}, Momentum space exciton composition of $\Omega_1$ and $\Omega_2$ have large band contribution along the $\Gamma-X$ direction.
	    \textbf{e}, Calculated macroscopic momentum-dependent dielectric function $\epsilon^{2d}(q_x,q_y)$ of freestanding 1L CrSBr exhibiting a strong anisotropy along $q_x$ and $q_y$.
	    \textbf{f}, $\epsilon^{2d}(0,q_y)$ and $\epsilon^{2d}(q_x,0)$ as a function of $q$. First-principles results are symbolized by the dots while solid lines are a guide to the eye.
	    \textbf{g}, Low-temperature ($\SI{4.2}{\kelvin}$) PL as a function of excitation power of bulk CrSBr. The spectrum shows a double resonance for the 1s exciton ($\sim\SI{1.366}{\electronvolt}$) due to the finite thickness of the flake ($\sim \SI{36.8}{\nano\meter}$) and two signatures $X^{*}$ and $X^{**}$ at $\SI{1.380}{\electronvolt}$ and $\SI{1.391}{\electronvolt}$, respectively. The energy splitting between the 1s exciton and $X^{*}$ is $\Delta E \sim \SI{15}{\milli\electronvolt}$.
     \textbf{h}, Reflectance contrast $\Delta R$ taken at $\SI{1.6}{\kelvin}$ showing the same resonances.
		\textbf{i}, Corresponding derivative $\Delta R / \Delta E$.
		\textbf{j}, False color plot of the magnetic field dependent differential reflectivity $\Delta R / \Delta E$ with the $B$-field applied parallel to the $c$-axis. The data show the 1s exciton, $X^{*}$ and a faint signature of the $X^{**}$.
		\textbf{k}, Position of the 1s exciton doublet and $X^{*}$. 
		}
\end{figure*}

The observed 1D nature of the electronic system in the 2D van der Waals material with 3D spin exchange interaction is unexpected. We can understand the behavior by considering the flat conduction band along the $\Gamma - X$ direction with its very large DOS (see Fig.~\ref{fig1}c). Under the assumption that the lattice excitations (phonons) couple to the highly anisotropic electronic continuum (collective excitations of electrons), we expect the BWF coupling term (between the electronic continuum and the discrete phonon line) to be proportional to the DOS at $E_F$. In the resonant Raman process, the excited phonons from the interband excitation can couple via intraband scattering processes to the continuum of electronic states that are in the lowest CB confined into a quasi-1D system. Using the 1D model for carrier dynamics (see SI), we can estimate this DOS to be $\nu(E) = \sqrt{\frac{2m^*}{E}}\frac{1}{2\pi \ell}$ where $m^*$ is the effective mass in the dispersive direction, $\ell$ is the effective interchain distance, and $E$ is the energy relative to the conduction band minimum. For carrier density $n_{\rm el}$, we find that this DOS scales as $\nu(E_F) \sim \frac{m^*}{\pi^2 \ell^2 n_{\rm el}}$, yielding a large number of electron-hole excitations which can hybridize with the phonon mode.\\

\textbf{Quasi-1D excitons with large binding energy.} CrSBr exhibits a rich optical spectrum with several peaks that are not yet identified.~\cite{Wilson.2021,Klein.2022,Bae.2022} The quasi-1D electronic structure of CrSBr is expected to directly affect the properties of the Coulomb-bound quasiparticle excitations. To evaluate the implications of the quasi-1D electronic structure on the complex optical spectrum, we calculate excitonic properties of monolayer CrSBr based on DFT+$GW$ and Bethe-Salpeter equations (BSE) and compare them with experimental signatures from optical PL and reflectivity measurements. Results for the calculated excitonic spectrum are shown for the electric field polarized along the $b$- and the $a$-directions (see Fig.~\ref{fig4}a and b). We identify four excitonic resonances in the $b$-direction: $\Omega_1 = \SI{1.22}{\electronvolt}$, $\Omega_2 = \SI{1.24}{\electronvolt}$, $\Omega_3 = \SI{1.87}{\electronvolt}$ and $\Omega_4 = \SI{1.93}{\electronvolt}$. The two energetically lowest excitons are split by only $\Delta E = \SI{20}{\milli\electronvolt}$. This energy difference has its origin in the two energy split conduction bands (see inset Fig.~\ref{fig1}c). We obtain large binding energies for $E_b (\Omega_1)$ = $\SI{0.88}{\electronvolt}$ and $E_b (\Omega_2)$ = $\SI{0.90}{\electronvolt}$. The corresponding real-space representation of the excitonic wavefunctions of $\Omega_1$ and $\Omega_2$ suggest a strong 1D character with the excitons extended along the $b$-direction but squeezed along the $a$-direction (see Fig.~\ref{fig4}c). The origin is the predominant orbital admixture from the flat conduction bands along the $\Gamma - X$ direction (see Fig.~\ref{fig4}d). For the most important orbitals we obtain an admixture for the upper conduction band of 61\% $d_{(x^2-y^2)}$, 10\% $d_{(3z^2-r^2)}$, 27\% $p_z$ and for the lower conduction band of 61\% $d_{(x^2-y^2)}$, 25\% $d_{(3z^2-r^2)}$, 7\% $p_z$. Importantly, the anisotropy can be found not only in the effective electron mass along $\Gamma - X $ and $\Gamma - Y $, but also dominates the dielectric function $\epsilon^{2D}(q_x,q_y)$ in reciprocal space (see Fig.~\ref{fig4}e and f) and is also manifested in the real and imaginary part of the macroscopic dielectric function in the monolayer and the bulk (see SI). It is this intricate interplay of effective mass anisotropy and dielectric anisotropy that shows in the quasi-1D character of the excitonic quasiparticles in CrSBr.

We now probe the excitonic signatures of bulk CrSBr at low temperature ($\SI{4.2}{\kelvin}$) by non-resonantly exciting the system with a continuous-wave laser at an energy of $\SI{2.384}{\electronvolt}$ (see Fig.~\ref{fig4}g). We observe that the 1s exciton at an energy of $\sim\SI{1.366}{\electronvolt}$ (at an excitation power of $\SI{10}{\micro\watt}$) appears as a doublet, likely due to a cavity effect (interference effects) caused by the thickness of the crystal ($\sim\SI{36.8}{\nano\meter}$) (see SI).~\cite{Dirnberger.2023} We obtain a very sharp Lorentzian peak with a full width at half maximum (FWHM) of $\sim\SI{1}{\milli\electronvolt}$ verifying high homogeneity of the parallel, weakly coupled chains, in combination with the high crystal quality. The 1s exciton red-shifts with increasing excitation power, as is typically observed for increasing number of photo-generated carriers in other semiconducting van der Waals materials (also see SI).~\cite{Steinhoff.2014}
Moreover, for high excitation powers, additional resonances above the 1s exciton, labelled $X^{*}$ and $X^{**}$, appear at energies of $\SI{1.380}{\electronvolt}$ and $\SI{1.391}{\electronvolt}$, respectively. While they are weakly visible in PL, their appearance in the low-temperature ($\SI{1.6}{\kelvin}$) differential reflectivity measurements and derivative (see Fig.~\ref{fig4}h and i) suggests finite oscillator strength and band related transitions. The energy splitting of the 1s exciton and the $X^{*}$ is only $\Delta E  \sim\SI{15}{\milli\electronvolt}$. 

\begin{figure*}
	\scalebox{\figurescale}{\includegraphics[width=1\linewidth]{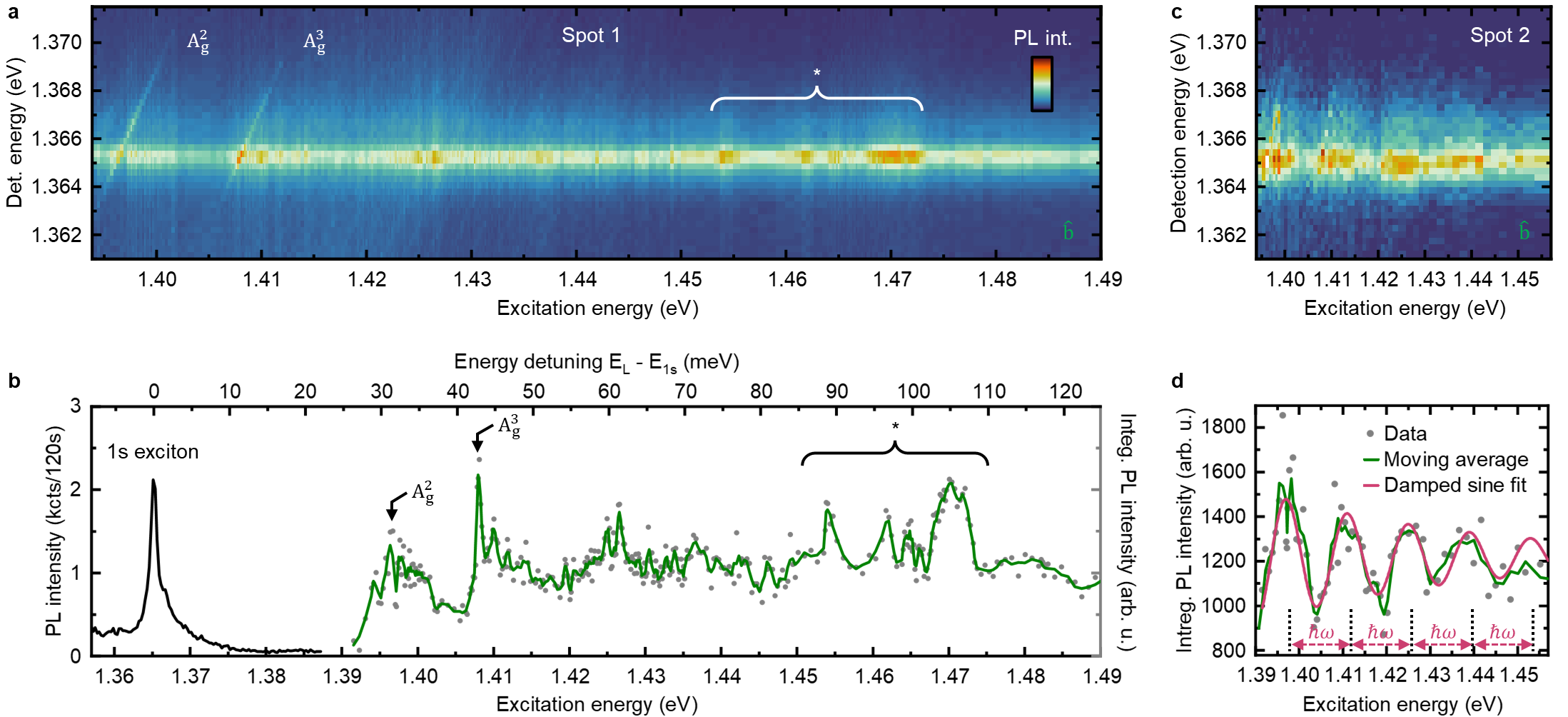}}
	\renewcommand{\figurename}{FIG.|}
	\caption{\label{fig5}
	    \textbf{Pronounced exciton-phonon coupling and rich electronic structure in bulk CrSBr.}
		\textbf{a}, False color low-temperature ($\SI{4.2}{\kelvin}$) PLE mapping of the 1s exciton at an emission energy of $\sim\SI{1.365}{\electronvolt}$. The 1s exciton shows resonant exciton-phonon coupling with two discrete phonons ($A_g^2$ and $A_g^3$) and additional electronic resonances highlighted with arrows. For the measurement an excitation power of $\SI{100}{\micro\watt}$ is used. \textbf{b}, PL emission spectrum of the 1s exciton at $E_L = \SI{1.406}{\electronvolt}$. Integrated intensity of the 1s exciton as a function of laser excitation energy. The green line is a moving average of three consecutive points to the data. The energies of the $A_g^2$ and $A_g^3$ are highlighted. At higher energies ($E >\SI{1.45}{\electronvolt}$) additional distinct energy resonances ($*$) are observed and highlighted.
        \textbf{c}, The PLE mapping at a second position with a lower energy resolution shows an oscillatory behavior in the intensity of the 1s exciton.
        \textbf{d}, Corresponding integrated intensity of the 1s exciton as a function of laser excitation energy. The green line is a moving average of three consecutive points to the data. The purple line is an exponentially damped $\sin$ fit to the data with a characteristic energy of $\hbar\omega = 13.99\pm 0.25\SI{}{\milli\electronvolt}$ (see dashed lines). For the measurement an excitation power of $\SI{100}{\micro\watt}$ is used.
		}
\end{figure*}

We next measure the impact and potential origin of the magnetic order and exchange interaction between the magnetic moments in the 3D matrix and the quasi-1D excitons by probing the 1s exciton, $X^{*}$ and $X^{**}$ at $\SI{1.6}{\kelvin}$ in reflectivity as a function of a magnetic field along the $c$-axis (see Fig.~\ref{fig4}j). The signatures show the expected magnetic field-dependent energy shift due to the change in magnetic ordering from AFM to FM until reaching the coercive field of $\SI{2}{\tesla}$ (see Fig.~\ref{fig4}j and k). The energy shift of the $X^{*}$ is weaker as compared to the 1s exciton. The $X^{**}$ is weak in signal for higher fields but suggests a qualitatively similar energy shift as the $X^{*}$ (see Fig.~\ref{fig4}j).

The attribution of the $X^{*}$ and $X^{**}$ is non-trivial. Potential scenarios would involve momentum direct or indirect transitions between the split conduction bands and the valence band. It is indeed a possibility that these transitions are more complex in nature with a momentum indirect character due to the strong extension of the wavefunction along the $\Gamma - X$ direction. Such a transition can furthermore be phonon-assisted. 

The energetic shift in the magneto-reflectivity data of the 1s exciton and the $X^{*}$ and $X^{**}$ suggests that the main orbital composition is likely still from the two split conduction bands. At the $\Gamma$ point in bulk CrSBr, the orbital composition of the lower conduction band is 59\% $d_{(x^2-y^2)}$, 21\% $d_{(3z^2-r^2)}$, 10\% $p_z$ while for the upper conduction band it is 60\% $d_{(x^2-y^2)}$, 5\% $d_{(3z^2-r^2)}$, 34\% $p_z$. The upper conduction band has a much lower admixture of Cr $d_{(3z^2-r^2)}$ orbitals which suggests a scenario in which the exciton reacts less to an external magnetic field since the magnetic moment is situated on the $d$-orbital. This would suggest the potential origin of the $X^{*}$ and $X^{**}$ from a transition that involves the upper conduction band. In general, the excitonic transitions are expected to inherit the orbital character, and this should also be the case for potential indirect transitions that preserve the orbital conduction band admixture away from the $\Gamma$ point. This further illustrates that the quasi-1D electronic structure and related orbital admixture have direct consequences on the magneto-optical coupling of the excitonic complexes in CrSBr.

\textbf{Pronounced exciton-phonon coupling and rich electronic structure.} In the last part of this investigation, we further confirm our interpretation of CrSBr as a quasi-1d material in the bulk limit by studying the electronic structure and quasiparticle interactions of CrSBr using high-resolution PLE measurements. We spectroscopically probe CrSBr at low temperature ($T = \SI{4.2}{\kelvin}$) by tuning a continuous-wave laser in small steps over a wide energy range from $\SI{1.39}{\electronvolt}$ to $\SI{1.76}{\electronvolt}$ and collect the PL from the 1s exciton with excitation and detection co-polarized along the $b$-direction. Figure~\ref{fig5}a shows a false color contour plot of the 1s exciton PL as a function of the excitation energy, with the corresponding integrated 1s exciton intensity shown in Fig.~\ref{fig5}b. The PLE measurement exhibits an intricate fine structure with several features and resonances. Two energetic regions are of particular interest. In the energy range ($E<\SI{1.45}{\electronvolt}$) close to the 1s exciton ($\sim\SI{1.365}{\electronvolt}$) phonons play an important role. Here, we observe two distinct lines crossing the 1s exciton PL signal. We interpret these lines as resonant Raman signatures suggesting pronounced exciton-phonon coupling between two discrete phonons with the 1s exciton. Such signatures appear in the spectra when the energy detuning of the laser photon and the 1s exciton equals a phonon or multiple phonon energies, leading to the resonant enhancement of the exciton emission. The two phonon lines cross with the 1s exciton at energy detunings of $\sim\SI{31.1}{\milli\electronvolt}$ ($\sim\SI{249}{\per\centi\meter}$) and $\sim\SI{42.9}{\milli\electronvolt}$ ($\sim\SI{343}{\per\centi\meter}$), respectively. The energies are in excellent agreement with the energies of the $A^2_g$ and $A^3_g$ phonon modes (see Fig.~\ref{fig3}). Pronounced exciton-phonon coupling effects are usually observed for 2D materials like hBN-encapsulated TMDCs~\cite{Chow.2017,Shree.2018} as a consequence of narrow exciton linewidths, intrinsically high oscillator strength and high phonon DOS $\propto q$~\cite{Shree.2018} but are very pronounced even in our measurement, further reflecting the low-dimensional character of CrSBr in the bulk limit.

In the same energy range ($E < \SI{1.45}{\electronvolt}$), we also observe oscillations in the 1s exciton emission intensity with equidistant energies with respect to the 1s exciton emission. These oscillations become even more apparent when collecting similar data on a second position but with a lower energy resolution to sample less fine structure (see Fig. \ref{fig5}c and d). The intensity of the 1s exciton shows an oscillatory intensity modulation with a characteristic energy of $13.99 \pm\SI{0.25}{\milli\electronvolt}$ ($112 \pm \SI{2}{\per\centi\meter}$) as obtained from an exponentially damped $\sin$ fit (see dashed lines in Fig.~\ref{fig5}d). This energy is in excellent agreement with the $A_g^1$ phonon mode energy ($\SI{14.3}{\milli\electronvolt}$) (see Fig.~\ref{fig3}). The observation of an optically dark phonon replica above the 1s exciton in energy has been observed in NiPS$_3$~\cite{Hwangbo.2021,Ergeen.2022} and interpreted as optically dark magnetic bound exciton-phonon states. A reduced energy of this bound state is expected.\cite{Toyozawa.1968,Hwangbo.2021} The observation of such signatures above the exciton in bulk CrSBr further illustrates the strong orbital and quasiparticle interactions that likely originate from the reduced dimensionality.

Finally, for higher energy detunings ($E>\SI{1.45}{\electronvolt}$), we observe a sequence of additional resonances in the 1s exciton emission that are close in energy to the single-particle gap (see peaks labelled with * in Fig.~\ref{fig5}a and b). In this energy range, Rydberg states can be expected in bulk CrSBr. In general, the anisotropic, quasi-1D screening is expected to result in a divergence of the 1s state, increasing the energy splitting between the 1s and 2s state. While our results provide initial insights into the complex electronic structure of CrSBr, additional measurements and theory are required to understand potential excited states in this material.

\section{Conclusions} 

We have provided strong experimental and theoretical evidence that bulk CrSBr is a quasi-1D material. The quasiparticles in CrSBr and their mutual couplings illustrate a 1D electronic character that offers very clean experimental signatures in addition to the magnetic order. We find that this material offers exciting opportunities to study the physics of quasi-1D electronic systems that are highly interacting with vibrational, optical and excitonic excitations and magnetic orderings of mixed dimensionalities. 

The energetic flat band along the $a$-direction ($\Gamma - X$ direction) motivates more detailed studies of potential correlated behaviors in this material. This is further suggested by our observation of a structural distortion in thin layers reminiscient of a Peierls instability. Generally, 1D systems are known to have strong tendencies towards charge density-wave order via Peierls instability, which is also consistent with the observation of pronounced electron-phonon interaction in this system. It is also realistic that at low densities other phases may compete and be favored. In particular, since the Coulomb interaction is long-ranged in the absence of free carriers, it is possible that at low carrier densities, the system may instead tend towards a 1D Wigner crystal~\cite{Schulz.1993} or a stripe phase,~\cite{Spivak.2004,Emery.2000} given that the kinetic energy is already largely quenched in 1D, while the Coulomb interaction remains long-range (see theoretical considerations in the SI).

CrSBr is a fascinating material that exhibits an intriguing electronic structure. We suggest that bulk CrSBr is best characterized as a stack of weakly coupled monolayers that each host excitons with large binding energies. The rich magnetic phase diagram contributes additional complexity rendering it an ideal platform for exploring fundamental physics and exciting applications. We anticipate that the 1D electronic character will prove to be a universal feature of the chalcogen-halide materials and signatures of this are likely to be expected in other related compounds. 

\section{Methods}

\subsection{Crystal growth and sample fabrication}

CrSBr bulk crystals were grown by chemical vapor transport.~\cite{Klein.2021} Samples were fabricated by mechanical exfoliation onto SiO$_2$/Si substrates. Sample thickness was verified by a combination of atomic force microscopy, optical phase contrast and Raman spectroscopy. For the electron microscopy studies, exfoliated flakes were transferred to TEM compatible sample grids using cellulose acetate butyrate (CAB) as polymer handle. Subsequently, the CAB was dissolved in acetone and the TEM grids rinsed in isopropanol prior to critical point drying.

\subsection{Raman spectroscopy}

Resonant and non-resonant Raman measurements were performed on a Renishaw inVia micro-Raman confocal microscope. For the measurements we used a 100x objective with a laser spot size of $\sim\SI{1}{\micro\meter}$. The laser excitation power used for the measurements taken at $\SI{785}{\nano\meter}$ and $\SI{532}{\nano\meter}$ was $\SI{1000}{\micro\watt}$ and $\SI{250}{\micro\watt}$, respectively. Ten individual spectra acquired at $\SI{20}{\second}$ were averaged resulting in a spectrum with a total integration time of $\SI{200}{\second}$. To adjust the polarization we used a linear polarizer in the excitation and detection path that we co-polarized. To change the polarization angle, the sample was placed on a rotating stage.

\subsection{Photoluminescence spectroscopy}

For the PL measurements, we mounted the sample in closed-cycle helium cryostats (Montana Instruments or AttoDry 800) with a base temperature of $\SI{4.2}{\kelvin}$. In both setups, measurements were made through the side window using a home-built confocal microscope with a ×100, 0.9 NA objective (Olympus). We excited the sample using a continuous-wave (CW) laser at $\SI{2.384}{\electronvolt}$ or $\SI{1.746}{\electronvolt}$. The PL is collected confocally after using a long-pass filter. For the polarization resolved PL, we used a linear polarizer in both excitation and detection and a half-wave plate to rotate the polarization. The light was fiber-coupled and directed to a high-resolution spectrometer that was attached to a liquid nitrogen cooled charge-coupled device camera.

\subsection{Photoluminescence excitation spectroscopy}

For the PLE measurements we used a tunable CW Ti:Sapphire laser (MSquared Solstis) that is highly monochromatic with a linewidth of $10^{-6}\SI{}{\nano\meter}$. The wavelength is locked using a wavemeter (High Finesse WS8). After wavelength tuning, we adjusted the excitation power to $\SI{100}{\micro\watt}$ before the microscope objective. Measurements were taken with the half-wave plate rotated such that the electric field pointed along the $b$-direction. The light was fiber-coupled and subsequently dispersed using a high-resolution spectrometer attached to a liquid nitrogen cooled charge-coupled device camera.

\subsection{Magneto-optical spectroscopy}

For the magnetic field-dependent optical measurements we mounted CrSBr bulk flakes on top of a $\textrm{SiO}_2/\textrm{Si}$ substrate into a closed-cycle cryostat (attodry 2100). We cooled the sample to a lattice temperature of around $\SI{1.6}{\kelvin}$. The magnetic field of a superconducting solenoid magnet is aligned along the $c$-axis of the crystal, with a maximum field up to $\SI{9}{\tesla}$. We fiber-coupled a broadband light source on the input to perform magnetic field-dependent reflectance measurements. The signal from the sample was collected using a second fiber and directed towards a high-resolution spectrometer attached to a liquid nitrogen-cooled charge-coupled device camera. The broadband laser excitation power was set to $\SI{100}{\micro\watt}$.

\subsection{Scanning tunneling microscopy}

The topographic images were taken at room temperature with a Unisoku UHV-LT four-probe scanning tunnelling microscope operated with a Nanonis controller, and equipped with a scanning electron microscope to allow precise location of scan areas. The CrSBr bulk crystal was cleaved under vacuum to obtain a clean sample surface. Using PtIr tips, we acquired differential conductance (dI/dV) spectra using a lock-in amplifier. During data acquisition, the feedback loop was switched off.  

\subsection{Scanning transmission electron microscopy}

STEM imaging was performed with a probe-corrected Thermo Fisher Scientific Themis Z G3 operated at $\SI{200}{\kilo\volt}$. The probe convergence semi-angle was 19 mrad. The corresponding probe size of the aberration-corrected electron beam is sub-Angstrom. Typical beam current used was $40-\SI{60}{\pico\ampere}$. All images were collected at room temperature. The frame size was 1024x1024 pixel and the dwell time is $\SI{500}{\nano\second}$/pixel. 10 images were acquired and overlapped using Velox software to increase the signal-to-noise ratio.

The STEM-HAADF images were analyzed using a custom-made Python code framework. The change in lattice distance between different atomic columns was detected by performing peak finding on the output of a fully convolutional neural network which predicts the atomic position as Gaussians in a segmentation map.

The percentage change between Cr and S/Br columns was calculated by $\Delta b = \frac{x - b_{mean}/2}{b_{mean}/2}$ where $x$ is the analyzed distance between two atomic positions and $b_{mean}/2$ the mean inter-atomic column value. The value was obtained from a 6L CrSBr that was used as a reference and extracted by fitting a histogram of inter-atomic column distances with a normal distribution.

\subsection{Many-body perturbation theory: $GW$+BSE}
CrSBr exhibits spin-orbit coupling and inplane ferromagnetism. 
To calculate its electronic ground state from first principles, we have employed density functional theory (DFT) in the generalized gradient approximation (GGA).~\cite{gga_pw}
The correct spin structure is provided by the noncollinear DFT formalism.~\cite{stark2011magnetic,sandratskii1998noncollinear}
This provides the wavefunction as spinors and is a good starting point for many-body perturbation theory, which straightly includes all effects of noncollinear magnetism as well.\\
On top of DFT, we have calculated the quasiparticle (QP) band structures by including the self energy $\Sigma(E)$ in the $GW$ approximation.~\cite{hedin1965new} Here, the one-particle Green’s function $G$ and the screened Coulomb interaction $W$ (including the dielectric response in random phase approximation) are derived and $\Sigma = iGW$ replaces the DFT exchange correlation energy $V_{\text{xc}}$.
The Hamiltonian becomes
\begin{equation}
H^{\text{QP}} = H^{\text{DFT}} + iG W - V_{\text{xc}}
\end{equation}
where the resulting difference is the quasiparticle correction $\Delta^{\text{QP}}$ from many-body perturbation theory. The QP band structure energies are then given by
\begin{equation}
E^{\text{QP}}_{n \mathbf{k}}=~E^{\text{DFT}}_{n \mathbf{k}} + \braket{\psi^{\text{DFT}}_{n \mathbf{k}} | \Sigma(E^{\text{QP}}_{n\mathbf{k}}) -V_{\text{xc}} | \psi^{\text{DFT}}_{n \mathbf{k}}} 
\end{equation}
While often the one-shot $GW$ is evaluated only accounting for the diagonal terms of $\Sigma$ (and $\Delta^\text{QP}$, respectively), this is not sufficient for CrSBr (see e.g. \cite{deilmann2020valley,forster2015two,heissenbuettel2021cri3}). Therefore, we calculate the QP wavefunctions as linear combination of the DFT wavefunctions
\begin{equation}
\psi^{\text{QP}}_{m \mathbf{k}} = \sum_n D^n_{m \mathbf{k}}  \psi^{\text{DFT}}_{n \mathbf{k}} ~.
\end{equation}
where the coefficients $D^n_{m\vec{k}}$ are given by the solution of the full eigenvalue problem
\begin{equation}
\sum_{n'} H^{\text{QP}}_{nn'} \left(\mathbf{k};E^{\text{QP}}_{m \mathbf{k}}\right) 
D^{n'}_{m \mathbf{k}} = E^{\text{QP}}_{m \mathbf{k}} D^{n}_{m \mathbf{k}}  ~. \label{eq:diagonalization} 
\end{equation}
For further details see.~\cite{heissenbuettel2021cri3}
Because of the large changes of the band gap, we use a scissor operator to anticipate its opening and thus accelerates the self-consistent determination of QP shifts. For monolayer, bilayer and bulk we employ $\SI{1.4}{\electronvolt}$, $\SI{1.1}{\electronvolt}$ and $\SI{1.0}{\electronvolt}$, respectively.\\
In $GW$ we represent all the two-point functions ($P,\epsilon,W$) by a hybrid basis set of Gaussian orbitals with decay constants ranging from 0.14 to 5.1\,$a_\text{B}^2$ and plane waves with an energy cutoff set to 1.5 Ry. We apply a $k$-point sampling of $17\times 13\times 1$ point in the first Brillouin zone. For the BSE we increased the sampling to $32\times 24\times 1$ points. For the resulting spectra (Fig.~\ref{fig4}a and b) an artificial broadening of $\SI{35}{\milli\electronvolt}$ is applied.

\subsection{Macroscopic momentum-dependent dielectric function}

To obtain the nonlocal macroscopic dielectric function in Fig~\ref{fig4} we calculate the head element of the dielectric matrix in the static limit $\varepsilon^{\mathrm{2d}}(\bq):=\varepsilon_{\mathbf{G}=0,\mathbf{G}'=0}(\bq)$ using Berkeley$GW$~\cite{PhysRevB.34.5390, deslippe_berkeleygw_2012} with 2-d Coulomb truncation.~\cite{PhysRevB.93.235435} To this end, spin-polarized density functional theory (DFT) calculations are carried out using \textsc{Quantum Espresso}~\cite{giannozzi_quantum_2009, giannozzi_advanced_2017}. We apply the generalized gradient approximation (GGA) by Perdew, Burke, and Ernzerhof (PBE) \cite{perdew_generalized_1996, perdew_generalized_1997} and use optimized norm-conserving Vanderbilt pseudopotential~\cite{van_setten_pseudodojo_2018} at a plane-wave cutoff of $80$~Ry. Uniform meshes with $24\times18\times1$ k-points are combined with a Fermi-Dirac smearing of $5$~mRy. 
 A large vacuum of 28\AA\,with a truncated Coulomb interaction~\cite{PhysRevB.96.075448} is included between repeated supercells in c-axis direction. The calculations are performed using the same lattice structure and band gap as in the electronic ground state calculation for our DFT+$GW$ approach. 
 
\subsection{Charge density calculations}

The charge density calculations shown in Fig.~\ref{fig1}g and h were performed using the PWmat package. SG15 pseudopotential and HSE exchange correlation functional were used. A plane wave energy for the basis set is cut off up to $50$ Ry. The charge density shown includes the states in the range $\SI{0.1}{\electronvolt}$ below the top of the valence band and $\SI{0.1}{\electronvolt}$ above the bottom of the conduction band.

%
%
\section{Acknowledgements}
J.K. and M.F. acknowledge support by the Alexander von Humboldt foundation. B.P. is a Marie Skłodowska-Curie fellow and acknowledges funding from the European Union’s Horizon 2020 research and innovation programme under the  Grant Agreement No. 840968 (COHESiV). F.M.R. acknowledges funding from the U.S. Department of Energy, Office of Basic Energy Sciences, Division of Materials Sciences and Engineering under Award DE‐SC0019336 for STEM characterization. Work by J.B.C. and P.N. is partially supported by the Quantum Science Center (QSC), a National Quantum Information Science Research Center of the U.S. Department of Energy (DOE). J.B.C. is an HQI Prize Postdoctoral Fellow and gratefully acknowledges support from the Harvard Quantum Initiative. Z.S. is supported through the Department of Energy BES QIS program on `Van der Waals Reprogrammable Quantum Simulator' under award number DE-SC0022277 for the work on long-range correlations, as well as partially supported by the Quantum Science Center (QSC), a National Quantum Information Science Research Center of the U.S. Department of Energy (DOE) on probing quantum matter.
P.N. acknowledges support as a Moore Inventor Fellow through Grant No. GBMF8048 and gratefully acknowledges support from the Gordon and Betty Moore Foundation as well as support from a NSF CAREER Award under Grant No. NSF-ECCS-1944085. M.F. and A.S. were supported through a grant for CPU time at the HLRN (Berlin/G\"ottingen). Z.Sof. and J.L. were supported by Czech Science Foundation (GACR No. 20-16124J). The optical PL and PLE spectroscopy work at low temperatures was supported by ARO MURI (Grant No. W911NF1810432), HEADS-QON (Grant No. DE-SC0020376), ONR MURI (Grant No. N00014-15-1-2761), CIQM (Grant No. DMR-1231319), and ONR (Grant No. N00014-20-1-2425). Work at CCNY was supported through the NSF QII TAQS (V.M.M.) and the DARPA Nascent Light Matter program (R.B.). F.D. was funded by the Deutsche Forschungsgemeinschaft (DFG, German Research Foundation) through Projektnummer 451072703.
M.H., T.D. and M.R. gratefully acknowledge the financial support from German Research Foundation (DFG Project No. DE 2749/2-1), the DFG Collaborative Research Center SFB 1083 (Project No. A13), and computing time granted by the John von Neumann Institute for Computing (NIC) and provided on the super-computer JUWELS at Jülich Supercomputing Centre (JSC). A.A., J.Q. acknowledge support from a Vannevar Bush Faculty Fellowship, AFOSR DURIP and MURI programs and the Simons Foundation. This work was carried out in part through the use of MIT.nano’s facilities.

\section{Author contributions}
J.K. and F.M.R. conceived the project and designed the experiments, J.K. and K.T. prepared the samples, J.K. and A.P. collected STEM data, J.K. and B.P. performed PL and PLE optical measurements, J.K. and K.T. performed polarization resolved Raman measurements, F.D., R.B. and J.Q., performed magneto-reflectivity measurements, R.D. and J.K. performed STM/STS measurements, Z.Sof. and J.L. synthesized high-quality CrSBr crystals, M.F., M.-C.H., A.S., T.D. and M.R. provided DFT+$GW$ and BSE calculations, J.K. and M.W. analyzed the experimental data, Z.S., provided charge density calculations with input from P.N. and furthermore provided the $k \cdot p$ model, J.B.C. provided additional theoretical insights into the correlated physics, M.L., U.W., V.M.M and A.A. discussed results, J.K. wrote the manuscript with input from all co-authors.\\

\section{Additional information}

\subsection{Supporting Information} Three band $k\cdot p$ model; Monolayer and bulk anisotropic dielectric function; STEM-HAADF of bulk to monolayer; Raman spectra of CrOCl; Raman selection rules; Laser excitation power dependence for different thicknesses; Power dependent exciton red-shift; Photoluminescence excitation spectroscopy of the 1s exciton; Oscillations in the photoluminescence signal of the 1s exciton; One-dimensional correlated phases; Interaction effects (charge density wave and stripe crystal). 

\subsection{Data availability} The data that support the findings of this study are available from the corresponding author on reasonable request.

\subsection{Code availability} The codes used for data analysis as well as \emph{ab initio} calculations are available from the corresponding author on reasonable request.

\subsection{Competing financial interests} The authors declare no competing financial interests.


%
%

\subsection{References}

\bibliographystyle{naturemag}
\bibliography{full}

\end{document}


\title{SI - The bulk van der Waals layered magnet CrSBr is a quasi-1D material}
%
\author{Julian~Klein}\email{jpklein@mit.edu}
\affiliation{Department of Materials Science and Engineering, Massachusetts Institute of Technology, Cambridge, Massachusetts 02139, USA}
\author{Benjamin~Pingault}
\affiliation{John A. Paulson School of Engineering and Applied Sciences, Harvard University, Cambridge, Massachusetts 02138, USA}
\affiliation{QuTech, Delft University of Technology, 2600 GA Delft, The Netherlands}
\author{Matthias~Florian}
\affiliation{Department of Electrical and Computer Engineering,  Department of Physics, University of Michigan, Ann Arbor, Michigan 48109, United States}
\author{Marie-Christin~Heißenbüttel}
\affiliation{Institut für Festkörpertheorie, Westfälische Wilhelms-Universität Münster, 48149 Münster, Germany}
\author{Alexander~Steinhoff}
\affiliation{Institut für Theoretische Physik, Universität Bremen, P.O. Box 330 440, 28334 Bremen, Germany}
\affiliation{Bremen Center for Computational Materials Science, University of Bremen,
28359, Bremen, Germany}
\author{Zhigang~Song}
\affiliation{John A. Paulson School of Engineering and Applied Sciences, Harvard University, Cambridge, Massachusetts 02138, USA}
\author{Kierstin~Torres}
\affiliation{Department of Materials Science and Engineering, Massachusetts Institute of Technology, Cambridge, Massachusetts 02139, USA}
\author{Florian~Dirnberger}
\affiliation{Department of Physics, City College of New York, New York, NY 10031, USA}
\author{Jonathan~B.~Curtis}
\affiliation{College of Letters and Science, UCLA, Los Angeles, CA 90095 USA}
\affiliation{John A. Paulson School of Engineering and Applied Sciences, Harvard University, Cambridge, Massachusetts 02138, USA}
\author{Mads~Weile}
\affiliation{Center for Visualizing Catalytic Processes (VISION), Department of Physics, Technical University of Denmark, DK-2800 Kgs. Lyngby, Denmark}
\author{Aubrey~Penn}
\affiliation{MIT.nano, Massachusetts Institute of Technology, Cambridge, Massachusetts 02139, USA}
\author{Thorsten~Deilmann}
\affiliation{Institut für Festkörpertheorie, Westfälische Wilhelms-Universität Münster, 48149 Münster, Germany}
\author{Rami~Dana}
\affiliation{Department of Materials Science and Engineering, Massachusetts Institute of Technology, Cambridge, Massachusetts 02139, USA}
\author{Rezlind~Bushati}
\affiliation{Department of Physics, City College of New York, New York, NY 10031, USA}
\affiliation{Department of Physics, The Graduate Center, City University of New York, New York, NY 10016, USA}
\author{Jiamin~Quan}
\affiliation{Photonics Initiative, CUNY Advanced Science Research Center, New York, NY, 10031, USA}
\affiliation{Physics Program, Graduate Center, City University of New York, New York, NY, 10026, USA}
\author{Jan~Luxa}
\affiliation{Department of Inorganic Chemistry, University of Chemistry and Technology Prague, Technická 5, 166 28 Prague 6, Czech Republic}
\author{Zdenek~Sofer}
\affiliation{Department of Inorganic Chemistry, University of Chemistry and Technology Prague, Technická 5, 166 28 Prague 6, Czech Republic}
\author{Andrea~Al\`{u}}
\affiliation{Photonics Initiative, CUNY Advanced Science Research Center, New York, NY, 10031, USA}
\affiliation{Physics Program, Graduate Center, City University of New York, New York, NY, 10026, USA}
\author{Vinod~M.~Menon}
\affiliation{Department of Physics, City College of New York, New York, NY 10031, USA}
\affiliation{Department of Physics, The Graduate Center, City University of New York, New York, NY 10016, USA}
\author{Ursula~Wurstbauer}
\affiliation{Institute of Physics and Center for Nanotechnology, University of Münster, 48149 Münster, Germany}
\author{Michael~Rohlfing}
\affiliation{Institut für Festkörpertheorie, Westfälische Wilhelms-Universität Münster, 48149 Münster, Germany}
\author{Prineha~Narang}\email{prineha@ucla.edu}
\affiliation{College of Letters and Science, UCLA, Los Angeles, CA 90095 USA}
\affiliation{John A. Paulson School of Engineering and Applied Sciences, Harvard University, Cambridge, Massachusetts 02138, USA}
\author{Marko~Lon\v{c}ar}\email{loncar@seas.harvard.edu}
\affiliation{John A. Paulson School of Engineering and Applied Sciences, Harvard University, Cambridge, Massachusetts 02138, USA}
\author{Frances~M.~Ross}\email{fmross@mit.edu}
\affiliation{Department of Materials Science and Engineering, Massachusetts Institute of Technology, Cambridge, Massachusetts 02139, USA}
%
%
%
\date{\today}
%

%
\maketitle
%
%

\tableofcontents

\newpage


\section{Three Band k$\cdot$p Model}

Monolayer CrSBr holds space group 59. The direct optical absorption in CrSBr is predominantly around the $\Gamma$-point. At the $\Gamma$-point, the valence band transforms as the $B_{3g}$ irreducible representation, and the first conduction band transforms as the $A_g$ irreducible representation. According to the symmetry, we build a three-band $k \cdot p$ model around the $\Gamma$ point in the space spanned by the basis set of ($\ket{B_{1u}}$,$\ket{A_{g}}$,$\ket{B_{3g}}$). The perturbation is kept up to the second order term. The resulting Hamiltonian is as follows
\begin{equation}
\hat{H} =
\begin{pmatrix}
E_2 + \frac{\hbar^2k^2}{2 m_e} & 0 & \frac{\hbar k_y p_y}{m_e}\\
0 & E_1 + \frac{\hbar^2 k^2}{2 m_e} + \frac{\hbar^2 k_y^2}{m_2^2} \frac{p_y^2}{E_2 - E_0} & 0\\
\frac{\hbar k_y p_y}{m_e} & 0 & E_0 + \frac{\hbar^2 k^2}{2 m_e} + \frac{\hbar^2 k_y^2}{m_2^2} \frac{p_y^2}{E_0 - E_2}~.
\end{pmatrix}
\label{eq:kp}
\end{equation}
Here $p_y = \bra{B_{3g}}{\hat{p_y}}\ket{B_{1u}}$. For the modelling we used an energy splitting in the conduction band of $\Delta_{E_2-E_1} = \SI{33}{\milli\electronvolt}$ and $p_y = -1.5 eVm_e/\hbar$. The $k \cdot p$ band structure is in good accordance to the calculated DFT-$GW$ band structure (see SI Fig.~\ref{SIfig3bandkp}a and b) and captures particularly the electronic anisotropy.

%
\begin{figure*}[ht]
\scalebox{\figurescale}{\includegraphics[width=1\linewidth]{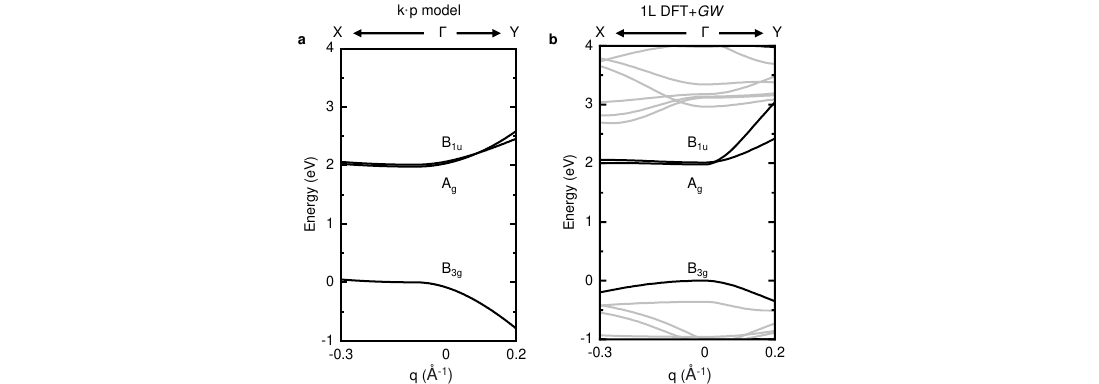}}
\renewcommand{\figurename}{SI Fig.|}
\caption{\label{SIfig3bandkp}
\textbf{Comparison between three band $k\cdot p$ model and DFT+$GW$ calculated band structure for monolayer CrSBr around the $\Gamma$-point.} 
\textbf{a}, Three band $k \cdot p$ band structure modelled around the $\Gamma$-point as obtained from SI Eq.~\ref{eq:kp}
\textbf{b}, Calculated DFT+$GW$ band structure of 1L CrSBr.
}
\end{figure*}
%
\newpage

\section{Monolayer and Bulk Anisotropic Dielectric Function}

%
\begin{figure*}[ht]
\scalebox{\figurescale}{\includegraphics[width=1\linewidth]{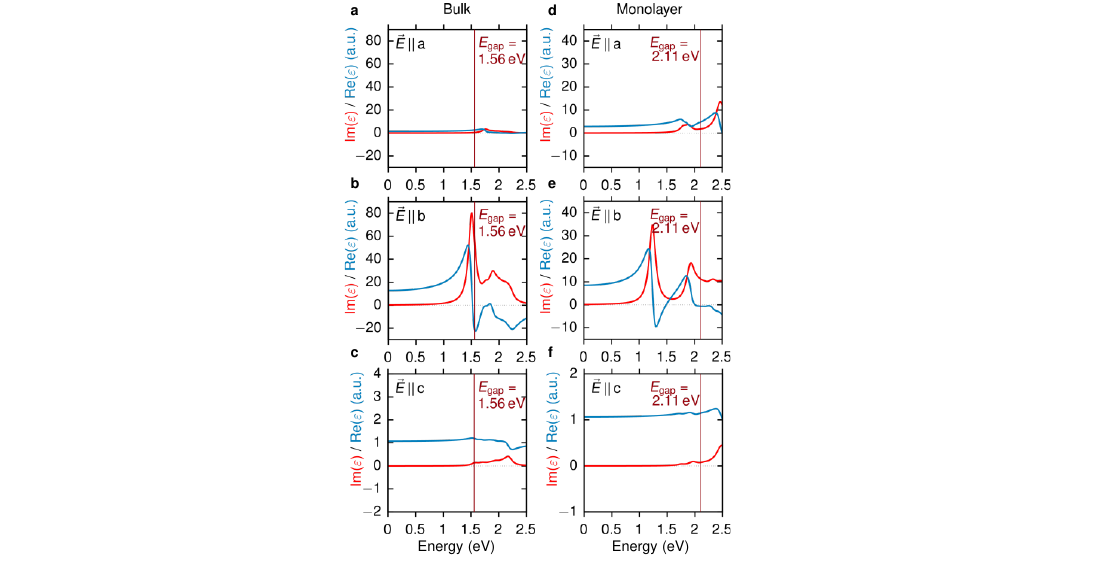}}
\renewcommand{\figurename}{SI Fig.|}
\caption{\label{SIfigepsilon}
\textbf{Real and Imaginary part of the macroscopic dielectric function of monolayer and bulk CrSBr.} 
By solving the Bethe-Salpether equation we find the real (blue) and imaginary part (red) of the dielectric function $\varepsilon$ which is connected to the reflection and absorption of the material. Panels \textbf{a} to \textbf{c} show the results for different directions of the polarization of the electric light field (inplane: a,b and out-of-plane: c) for bulk and \textbf{d} to \textbf{f} for monolayer. As the primitive unit cell of the bulk consists of two monolayers, the scale of $\varepsilon$ is adjusted accordingly. The largest response is obtained for light polarized parallel to the b axis (easy axis), while for light vertically to the van der Waals stacked layers the absorption nearly vanishes. The vertical line (dark red) represents the minimal direct gap (at Z in bulk and at $\Gamma$ in monolayer CrSBr). For the calculations we have applied a $k$-space grid of (20x27x3) in bulk and (24x32x1) in the monolayer.
}
\end{figure*}
%

\newpage

\section{STEM-HAADF of Bulk to Monolayer}

%
\begin{figure*}[ht]
\scalebox{\figurescale}{\includegraphics[width=1\linewidth]{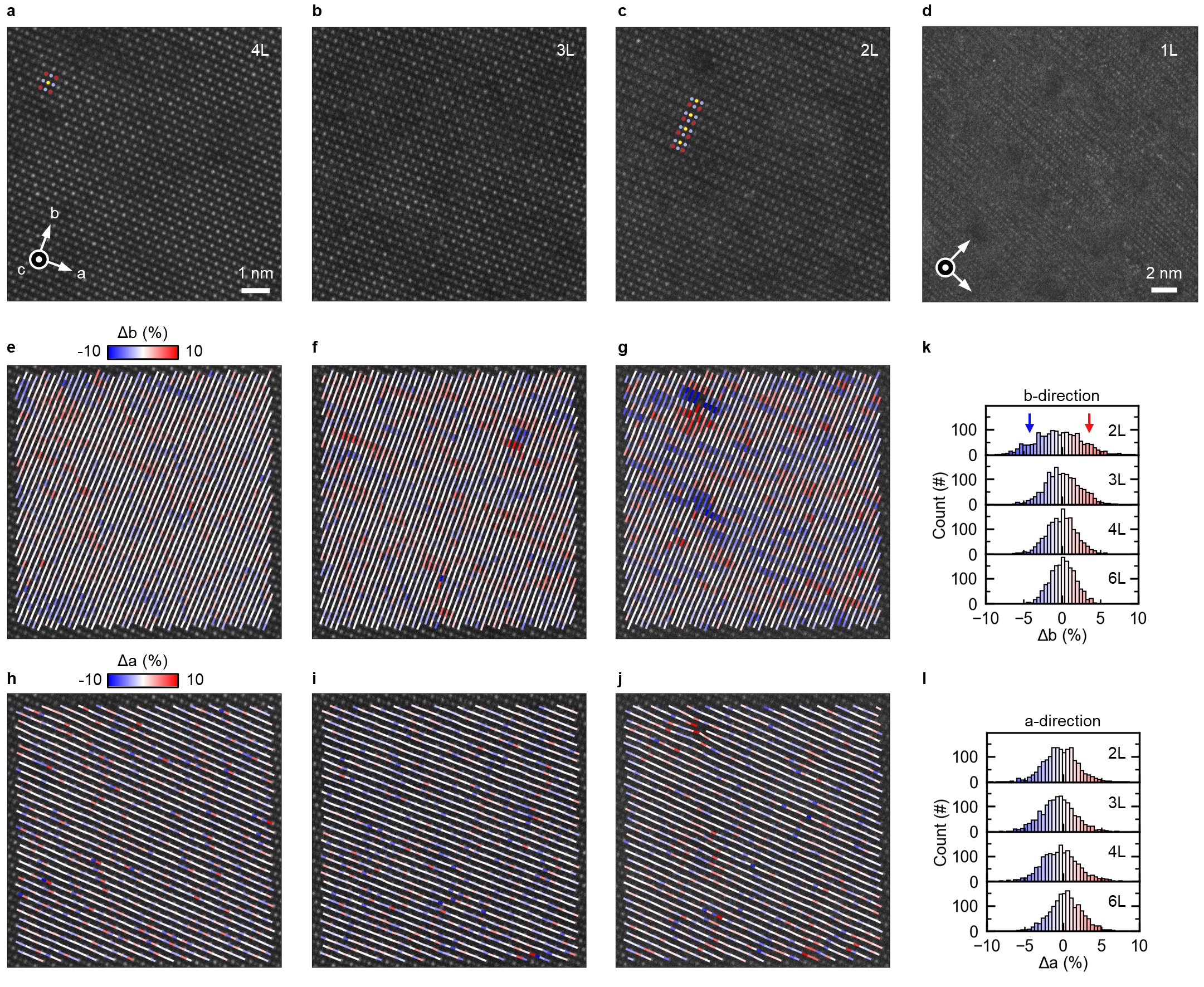}}
\renewcommand{\figurename}{SI Fig.|}
\caption{\label{SIfig3bandkp}
\textbf{STEM-HAADF of mono- and fewlayer CrSBr.} 
\textbf{a-d}, STEM-HAADF image of 4L, 3L, 2L and 1L CrSBr. The 4L, 3L and 2L images are collected at an electron beam energy of $\SI{200}{\kilo\electronvolt}$ and the 1L image at $\SI{300}{\kilo\electronvolt}$. A typical beam current of $\SI{40}{\pico\ampere}$ was used. The lattice temperature was $\SI{300}{\kelvin}$.
\textbf{e-g}, Corresponding map of the lattice distance change $\Delta b$ along the $b$-direction.
\textbf{h-j}, Corresponding map of the lattice distance change $\Delta a$ along the $a$-direction.
\textbf{k}, Layer dependent inter-atomic lattice distance histograms along the $b$-direction.
\textbf{l}, Layer dependent inter-atomic lattice distance histograms along the $a$-direction.
}
\end{figure*}
%

\newpage

\section{Raman Spectra of CrOCl.}

%
\begin{figure*}[ht]
\scalebox{\figurescale}{\includegraphics[width=1\linewidth]{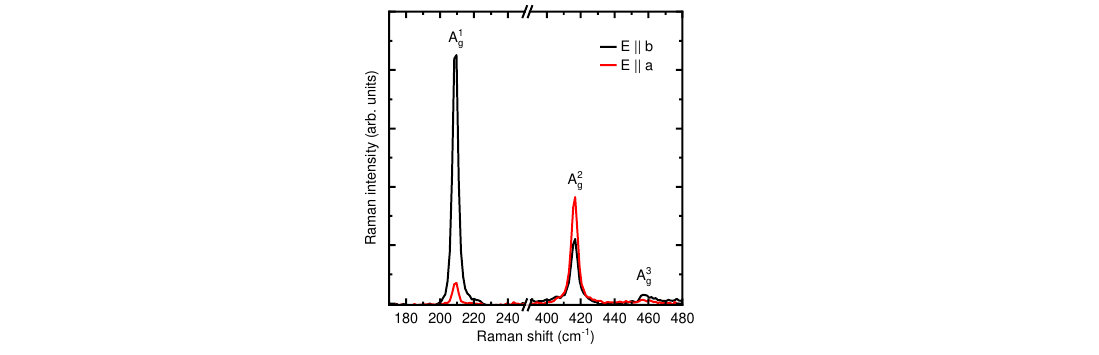}}
\renewcommand{\figurename}{SI Fig.|}
\caption{\label{SICrOCl}
\textbf{Raman spectra of bulk CrOCl.} 
Polarization dependent Raman spectra of bulk CrOCl for co-aligned and linearly polarized excitation and detection laser measured along the $a$- and $b$-direction. An excitation laser energy of $\SI{1.58}{\electronvolt}$ is used and an excitation power of $\SI{500}{\micro\watt}$. Similar to CrSBr, CrOCl shows the A$^1_g$, A$^2_g$ and A$^3_g$ modes. Moreover, the A$^1_g$ and A$^3_g$ mode show an intensity maximum along the $b$-direction, while the A$^3_g$ shows a maximum intensity along the $a$-direction in agreement with Ref.~\cite{Zhang.2019} Most importantly, the lineshape of all modes is symmetric and purely Lorentzian unlike for CrSBr for the same excitation energy due to the absence of resonance conditions in CrOCl owing to the sub-gap excitation.
}
\end{figure*}
%

\newpage

\section{Raman Selection Rules}

The Raman spectrum of pristine CrSBr shows three main phonon modes in the absence of resonant excitation which we identify as the A$^1_{g}$, A$^2_{g}$, and A$^3_{g}$ phonon modes.~\cite{Torres.2023} The crystal structure of CrSBr is orthorhombic with a Pmmn space group and a D$_{2h}$ point group. The unit cell of CrSBr contains in total 6 atoms and thus we expect 18 phonon modes in the first Brillouin zone. The irreducible representation of CrSBr at the $\Gamma$ point of the Brillouin zone is

\begin{equation}
    \Gamma = 3 A_{g} \otimes 2 B_{1u} \otimes 3 B_{2g} \otimes 2 B_{2u} \otimes 3 B_{3g} \otimes 2 B_{3u}  ~.
    \label{eq:irr_rep}
\end{equation}

We only include 15 phonon modes, as the first B$_{1u}$, B$_{2u}$, and B$_{3u}$ modes are acoustic with a zero frequency at the $\Gamma$ point. The A$_{g}$, B$_{2g}$, and B$_{3g}$ modes are Raman active with the corresponding Raman tensors for a Pmmn space group symmetry.~\cite{Loudon.1964}

\begin{equation}
    \mathcal{R}( A_{g}) = 
     \begin{pmatrix}
a &  &   \\
 & b &   \\
 &  & c  \\
\end{pmatrix}~,~\nonumber
\end{equation}

\begin{equation}
    \mathcal{R}( B_{2g}) = 
     \begin{pmatrix}
 &  & e   \\
 & &  &  \\
 e & & \\
\end{pmatrix}
 ~,~\mathcal{R}( B_{3g}) = 
     \begin{pmatrix}
 & & &   \\
 & & f   \\
 & f &   \\ 
\end{pmatrix} ~.~
    \label{eq:R_B}
\end{equation}

Based on the Raman tensor, only the three A$_{g}$ modes are expected in first order Raman scattering. The inelastically scattered light intensity $\mathcal{I}$ of each phonon mode in Raman scattering is determined by the following equation

\begin{equation}
   \mathcal{I} \propto | {e^{s}} \cdot \mathcal{R} \cdot {e^{i}}|^{2}
   ~,
    \label{eq:Raman_intensity}
\end{equation}

where ${e^{s}}$ and ${e^{i}}$ represent the vector of the scattered and incident photons, respectively. 





In this work, Raman experiments are conducted in a backscattering configuration with samples in the a-b-plane and a co-linearly polarized excitation laser. Thus the tensor that yields a non-zero intensity is the A$_{g}$ as shown below

\begin{equation}
   \mathcal{I} \propto | {a} \cdot {e^{s}_{x}e^{i}_{x}} + {b} \cdot {e^{s}_{y}e^{i}_{y}}|^{2}
   ~,
    \label{eq:Raman_intensity}
\end{equation}

While all A$_{g}$ modes are out-of-plane modes the intensity anisotropy differs between the modes, with the A$^1_{g}$ and A$^3_{g}$ showing the intensity maximum along the $b$-direction and the A$^2_{g}$ mode along the $a$-direction. This difference between the modes is not directly apparent just from the Raman tensors and might have its origin in the electronic structure of the material.

\newpage

\section{Laser Excitation Power Dependence for Different Thicknesses.}

%
\begin{figure*}[ht]
\scalebox{\figurescale}{\includegraphics[width=1\linewidth]{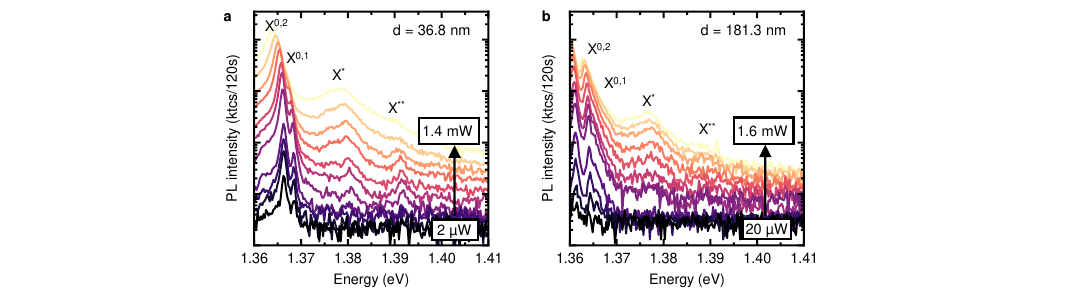}}
\renewcommand{\figurename}{SI Fig.|}
\caption{\label{SIfigCrystal_Structure}
\textbf{Excitation power dependent PL of bulk CrSBr of different thickness.} 
\textbf{a}, Power dependence of bulk CrSBr with a thickness of $\SI{36.8}{\nano\meter}$. The PL exhibits the 1s exciton that shows a double peak ($X^{0,1}$ and $X^{0,2}$) and two additional resonances $X^*$ and $X^{**}$.
\textbf{b}, Power dependence of bulk CrSBr with a thickness of $\SI{181.3}{\nano\meter}$. The 1s exciton is split into three peaks, likely due to interference effects from the finite thickness of the material. However, the resonances $X^*$ and $X^{**}$ are unchanged and appear at the approximately same energy difference with respect to the 1s exciton.
}
\end{figure*}
%


\section{Power Dependent Exciton Red-Shift}

%
\begin{figure*}[ht]
\scalebox{\figurescale}{\includegraphics[width=1\linewidth]{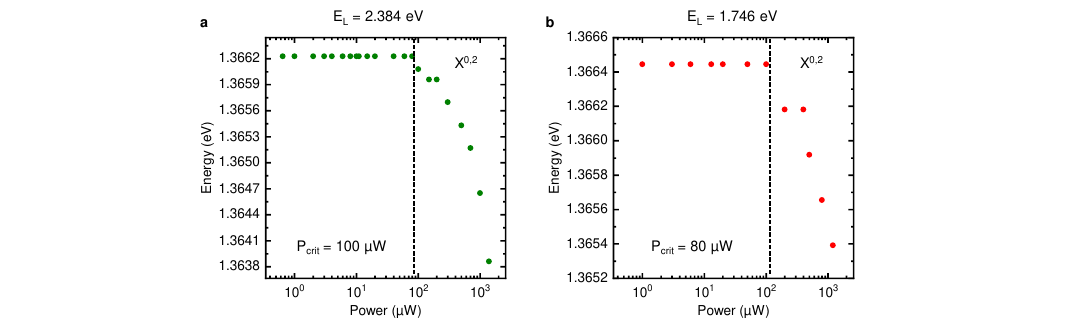}}
\renewcommand{\figurename}{SI Fig.|}
\caption{\label{SIfigCrystal_Structure}
\textbf{Power dependent exciton red-shift.} 
\textbf{a}, Exciton peak position as a function of laser excitation power for an excitation energy of $\SI{2.384}{\electronvolt}$. The exciton red-shifts at a critical power of $P_{crit} \sim \SI{100}{\micro\electronvolt}$.
\textbf{b}, Exciton peak position as a function of laser excitation power for an excitation energy of $\SI{1.746}{\electronvolt}$. The exciton red-shifts at a critical power of $P_{crit} \sim \SI{80}{\micro\electronvolt}$.
}
\end{figure*}
%

\newpage

\section{Photoluminescence Excitation Spectroscopy of the 1s Exciton.}

%
\begin{figure*}[ht]
\scalebox{\figurescale}{\includegraphics[width=1\linewidth]{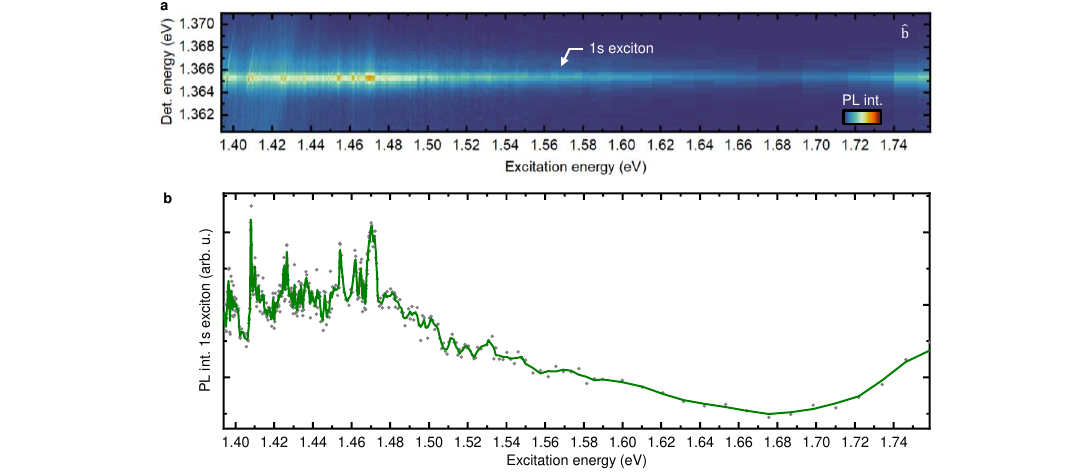}}
\renewcommand{\figurename}{SI Fig.|}
\caption{\label{SIfigPLE}
\textbf{Excitation laser energy dependent intensity of the 1s exciton of bulk CrSBr.} 
\textbf{a}, Spectrally integrated PL intensity of the 1s exciton for the laser energy tuned from $\SI{1.39}{\electronvolt}$ to $\SI{1.76}{\electronvolt}$. The solid line is a moving average to the data.
\textbf{b}, Spectrally integrated intensity of the 1s exciton emission. The solid green line is a moving average to the data. The intensity decreases until reaching a minimum at $\sim\SI{1.68}{\electronvolt}$ and increases towards higher energies due to contributions from higher lying bands.
}
\end{figure*}
%

\newpage

\section{Oscillations in the Photoluminescence Signal of the 1s Exciton.}

%
\begin{figure*}[ht]
\scalebox{\figurescale}{\includegraphics[width=1\linewidth]{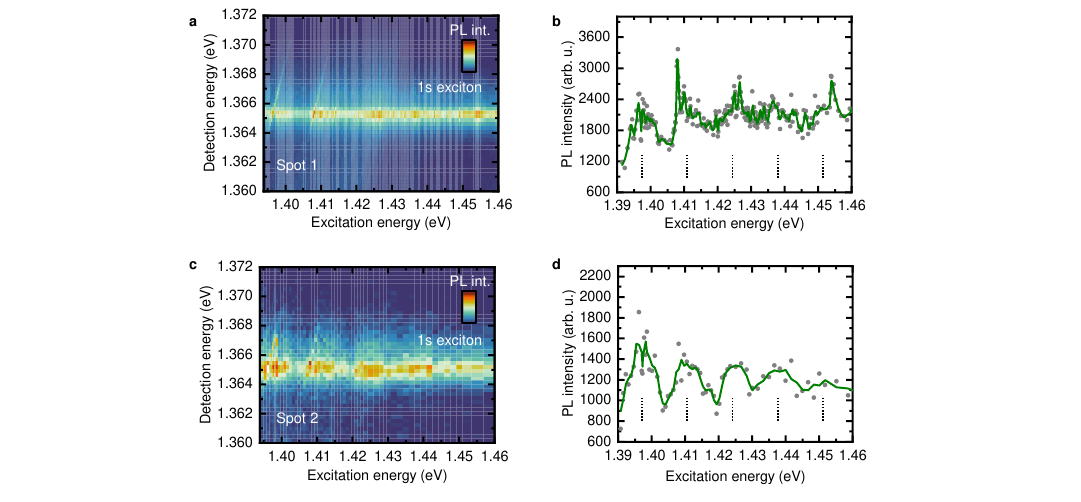}}
\renewcommand{\figurename}{SI Fig.|}
\caption{\label{SIfigPLE_pos2}
\textbf{Photoluminescence excitation spectroscopy of the 1s exciton in bulk CrSBr.} 
\textbf{a}, False color mapping of the excitation laser energy dependent spectral response of the 1s exciton at a sample temperature of $\SI{4.2}{\kelvin}$. The laser energy is tuned from $\SI{1.39}{\electronvolt}$ to $\SI{1.46}{\electronvolt}$ with fine energy steps. The 1s exciton exhibits intensity oscillations with equidistant energy spacing.
\textbf{b}, Spectrally integrated intensity of the 1s exciton emission. The solid line is a moving average to the data. The order of the oscillations is highlighted.
\textbf{c}, False color mapping of the excitation laser energy dependent spectral response of the 1s exciton taken at a second spot on the sample with larger energy steps.
\textbf{d}, Spectrally integrated intensity of the 1s exciton emission. The solid line is a moving average to the data. The approximate position of the oscillations is highlighted.
}
\end{figure*}

\newpage

\section{One-Dimensional Correlated Phases}

We provide additional insights into the potential correlated physics of CrSBr. Below the Curie temperature, this CrSBr exhibits intra-layer ferromagnetic order due to Stoner-type interactions resulting in a spin-polarized band structure. As we described in the main manuscript, the conduction band is highly anisotropic, and is nearly completely flat along one axis ($\Gamma - X$), and dispersive along the other ($\Gamma - Y$). This leads us to a model of the system at low carrier doping of that of weakly coupled electronic chains. The chains are furthermore spin-polarized due to the magnetic order (below the Curie temperature).

We consider a toy model Hamiltonian for the band structure of 
\begin{equation}
H = \sum_{\bf p} c_{\bf p}^\dagger E_{\bf p} c_{\bf p}
\end{equation}
where $c_{\bf p}^\dagger$ creates an electron in the conduction band with momentum $\bf p$.
Because of the large exchange energy, we neglect for the time being the spin minority band, though this may be important once fluctuations in the magnetic order are incorporated. 

The dispersion is modeled as 
\begin{equation}
E_{\bf p} = \frac{p_x^2}{2m^*} + t_{\perp}(1- \cos p_y \ell ),
\end{equation}
where $m^*$ is the effective mass of the conduction band along the dispersive direction, and $t_{\perp}$ is the bandwidth of the interchain tunneling, and $\ell$ is the interchain separation, on the order of one lattice constant.

We take $t_{\perp} \ell^2  \ll 1/ m^*$ so that the band dispersion is highly anisotropic. 
In the absence of electron interactions, this is already interesting as it implies the possibility for very strong impurity RKKY interactions and other one-dimensional electronic liquid physics. 
We begin by determining the density of states.

We have the density of states of the conduction band of 
\begin{equation}
\nu(E) = \int_{\bf p} \delta( E - E_{\bf p}) \sim \frac{1}{\ell} \int \frac{dp_x}{2\pi} \delta(E - p_x^2/2m^*)= \frac{2m^*}{\ell} \frac{2}{2\pi 2\sqrt{2m^* E}} = \sqrt{2m^*/E}\frac{1}{2\pi \ell},
\end{equation}
evaluated in the flat band limit. 
We see the emergence of a prominent van Hove singularity due to the low dimensionality.
We also see the factor of $1/\ell$ which is the effective density of chains in the $Y$-direction ($b$-direction), and is a very large factor owing to the lattice scale origin. 

Based on simple integration of density of states, or in general Luttinger's theorem, we find the area of the filled Fermi surface to be equal to the carrier density 
\begin{equation}
2 \pi/\ell \times 2k_F = 4\pi^2 n_{\rm el}
\end{equation}
in the case of a spin-polarized system. 
We have as a result the relation 
\begin{equation}
k_F = \pi \ell n_{\rm el} 
\end{equation}
and the Fermi level of 
\begin{equation}
E_F = \frac{\pi^2 \ell^2 n_{\rm el}^2}{2m^*}.
\end{equation}
We find the density of states at the Fermi level is 
\begin{equation}
\nu_F = \frac{m^*}{\pi^2 \ell^2 n_{\rm el} } .
\end{equation}
For purposes of estimation, we use the calculated effective mass $m^e_Y = 7.31 m_0$, and take an interchain separation of $\ell = 4\si{\angstrom}$.

Since the carrier density is likely to be small in comparison to the chain separation $\ell^2$, we expect a very small Fermi level even for appreciable doping. 
As a result, the interaction effects are likely to be dominated by electron-electron interactions and strong Coulomb energy. 

\section{Interaction Effects}
We now briefly discuss interaction effects.
Since the model is effectively single channel due to the ferromagnetic ordering, we have comparatively few possible outcomes.
The first is the standard Peierl's instability, resulting in a charge density wave (CDW) order, with density (and possibly therefore also magnetization) oscillations along the dispersive direction. 
We also briefly discuss the possibility of a second, related instability towards Wigner crystal order at low densities. 
Finally, we will discuss possible effects due to disorder, which is likely to be highly relevant in the low-dimensional transport. 

\subsection{Charge Density Wave}
For charge density wave order, we consider a model of the free electrons interacting with a phonon mode $b_{\bf q}$ via 
\begin{equation}
H_{\rm int} = \sum_{\bf p,q} \frac{g_{\bf p}({\bf q}) }{\sqrt{\rm Vol.}} \frac{b_{\bf q} + b_{-\bf q}^\dagger}{\sqrt{2\Omega_{\bf q}}} c_{\bf p+\frac{q}{2}}^\dagger c_{\bf p- \frac{q}{2}} + \sum_{\bf q} \Omega_{\bf q} b_{\bf q}^\dagger b_{\bf q}
\end{equation}
where the phonon mode is assumed to be of appropriate symmetry and dispersion, captured by $\Omega_{\bf q}$. 
We treat this in a perturbative manner, integrating out the fermionic operators. 

We find the phonon propagator is dressed by hybridization with the electron-hole excitations. 
This results in phonon Green's function, in equilibrium, of 
\begin{equation}
D(q) = \frac{2\Omega_{\bf q}}{(i\omega_m)^2 - \Omega_{\bf q}^2 -\int_p  |{g}_{\bf p}({\bf q})|^2 \mathscr{G}_0(p+\frac{q}{2})\mathscr{G}_0(p-\frac{q}{2})} .
\end{equation}
The ionic effective mass $M_{\rm eff}$ has been absorbed into the coupling constant $g$. 
 
We carry out the Matsubara sum and analytically continue to obtain 
\begin{equation}
D^R(\omega,\mathbf{q}) = \frac{2 \Omega_{\bf q}}{(\omega + i0)^2 - \Omega_{\bf q}^2  + \int_{\bf p} |g_{\bf p}(\mathbf{q})|^2 \frac{ f(\beta \xi_{\bf p + \frac{q}{2}}) - f(\beta \xi_{\bf p - \frac{q}{2}})}{  \xi_{\bf p + \frac{q}{2}} -  \xi_{\bf p - \frac{q}{2}}  - (\omega + i0^+)}}.
\end{equation}
Here, $f(z) = \frac12 \tanh(z/2)$ is the Fermi occupation function. 
To proceed, we need to specify more details about the phonon coupling function, and also numerically integrate the Lindhard function, in general. 

We focus on the static response function for the phonons, as this will signal the onset of the Peierl's phase. 
We have instability when 
\begin{equation}
\Omega_{\bf q}^2 =  \int_{\bf p} |g_{\bf p}(\mathbf{q})|^2 \frac{ f(\beta \xi_{\bf p + \frac{q}{2}}) - f(\beta \xi_{\bf p - \frac{q}{2}})}{  \xi_{\bf p + \frac{q}{2}} -  \xi_{\bf p - \frac{q}{2}}  }.
\end{equation}
In the case of a Holstein-type coupling to the density alone, we get $g_{\bf p}({\bf q}) = g({\bf q})$ and thus find the phase boundary determined by 
\begin{equation}
\Omega_{\bf q}^2/|g({\bf q})|^2 =\chi_0({\bf q}).
\end{equation}
where here we have introduced the bare static Lindhard function 
\begin{equation}
 \chi_0({\bf q}) = \int_{\bf p} \frac{ f(\beta \xi_{\bf p + \frac{q}{2}}) - f(\beta \xi_{\bf p - \frac{q}{2}})}{  \xi_{\bf p + \frac{q}{2}} -  \xi_{\bf p - \frac{q}{2}}  }.
\end{equation}

The key observation is that at zero temperature there is a Kohn anomaly in the density-density response due to the perfect nesting with the phonon mode exhibiting extreme softening at $q = 2k_F$ before onset of the CDW transition.
We have the Lindhard response at $T = 0$ of (for $q < 2k_F$)
\begin{equation}
\chi_0(q) = \frac{m^* }{\pi \ell |q_x| }\log\left( \frac{2k_F+ |q_x|}{2k_F - |q_x|}\right).
\end{equation}
Note this only depends on $q_x$ in the extreme flat band limit of $t_{\perp} \to 0$. 
This has a logarithmic singularity at the nesting wavevector of $2k_F$, which is the origin of the instability. 
We find as $q\to 2k_F$ from below that $\chi_0(2k_F) \to \frac{m}{\pi \ell 2k_F }\log(2/(1-q/(2k_F))) \to + \infty$. 
Therefore, the equation $ \Omega_{2k_F}^2/|g(2k_F)|^2 =\chi_0(2k_F)$ always has as solution at zero temperature.
We also see static screening of $\chi_0(q\to 0) = \nu(E_F)$ which is consistent with a Thomas-Fermi result. 

The complication is that CDW order requires temperature $T \ll E_F$. 
However, we find that $E_F$ also evolves with doping such that we need to solve this more carefully in the $T,n_{\rm el}$ plane. 
Note that $n_{\rm el}$ is related to the Fermi momentum by 
\begin{equation}
k_F = \pi \ell n_{\rm el}
\end{equation}
so that the density directly sets the wavelength of the CDW. 
We plot the function 
\begin{equation}
\chi_0( q = 2k_F, T )  = \frac{m^*}{\pi^2 \ell^2 n_{\rm el} } \int_{-\infty}^{\infty} \frac{dx}{2x} \left( \tanh\left( \frac{ ( x + 1)^2 - 1}{ 4 m^* T/ (\pi \ell n_{\rm el})^2 } \right) - \tanh\left( \frac{ ( x - 1)^2 - 1}{ 4 m^* T/ (\pi \ell n_{\rm el})^2 } \right)  \right),
\end{equation}
which depends on temperature relative to the Fermi level, and carrier density which in turn tunes the Fermi level. 
Note that in the tanh functions we ignore the $x^2$ term, which amounts to implementing the linearized Fermi surface. 


%
\begin{figure*}[ht]
\scalebox{\figurescale}{\includegraphics[width=0.5\linewidth]{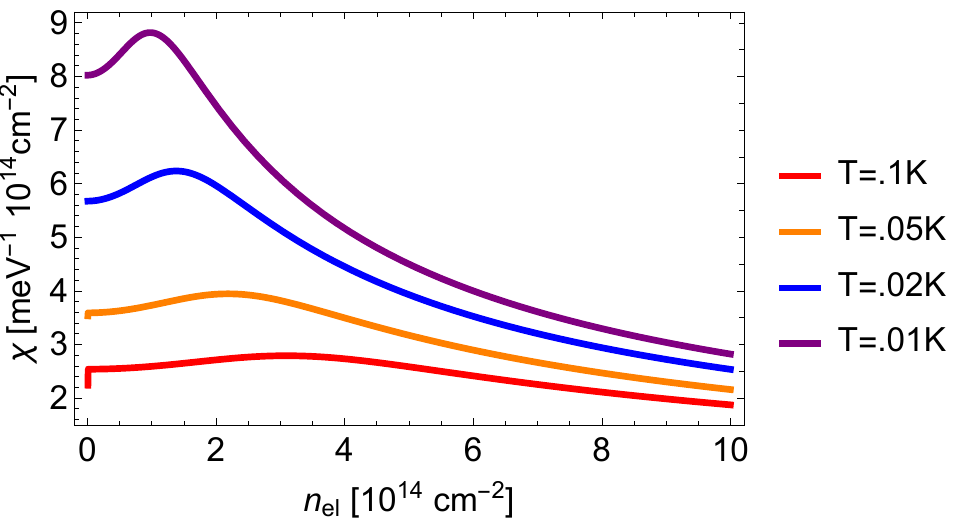}}
\renewcommand{\figurename}{SI Fig.|}
\caption{\label{fig:cdw-phase}
\textbf{Lindhard function.} 
Lindhard function at $q=2k_F$ as a function of carrier density evolving with temperature. The peak of this function determines if there is a charge-density wave instability, and at lower temperatures this peak becomes more pronounced, becoming singular only at zero temperature. As the carrier density increases this rises as the Fermi energy grows and eventually passes the temperature, establishing quantum coherence.
However, the density of states also falls off as $n_{\rm el}$ increases, with $\chi(2k_F)\sim 1/n_{\rm el}$.
Therefore, the potential emergence of a CDW phase clearly requires low temperatures and a high, but carefully tuned carrier density.
}
\end{figure*}
%

We ought to carry out a more systematic analysis with better estimates for parameters, but this is a decent starting point to guide the qualitative discussion. 
We now move to a related scenario, which involves the long-range Coulomb interaction rather than the phonon mediated interaction. 

\subsection{Stripe Crystal}
It is reasonable to expect a form of Wigner crystal to be stabilized.
In order to capitalize on both Coulomb repulsion and kinetic energy, a reasonable guess is a form of striped state where electron density occupies conducting chains but the filled chains space out so as to reduce Coulomb interaction.
Consider a system where the wavefunction is given by ansatz consisting of a product state across different chains, and within each chain a Fermi surface.
Let us consider each filled chain to be filled to Fermi level $k_F$ such that each chain carries a linear charge density, guaranteed by Luttinger's theorem of 
\begin{equation}
n_{\rm 1D} = k_F/\pi.
\end{equation}
This leads to an average kinetic energy per particle of 
\begin{equation}
\langle T\rangle = \frac{1}{n_{\rm 1D}} \int_{-k_F}^{k_F} \frac{dp}{2\pi} \frac{p^2}{2m} = \frac{1}{\pi n_{\rm 1D}}\frac{k^3_F}{6m} = \frac{E_F}{3 }.
\end{equation}

Now, let us consider the equal spaced chains with distance $b$ between them.
Then we consider an ansatz of every $\nu=1,2,...$ chains being filled (so for $\nu = 1$ we have every chain filled, for $\nu = 2$ every other, so on), with the other chains being empty. 
This gives an overall two-dimensional charge density of 
\begin{equation}
n_{\rm 2D} = n_{\rm 1D}/(\nu b) = \frac{k_F}{\nu \pi b}.
\end{equation}

We now take the Coulomb repulsion into account. 
We use a rough model which only accounts for the repulsion between neighboring chains.
We also use a model based on a point charge, and it is unclear at the moment how to generalize this to continuum distributions of lines of uniform charge.
Nevertheless, we know the potential due to a linear distribution of charges when the charge is distance $r$ away
\[
V = - \frac{e n_{\rm 1D}}{2\pi \epsilon_0}\log(r/b) 
\]
with a cutoff placed on the logarithm of order $b$, the lattice constant. 
The energy per unit area is 
\begin{equation}
U = -\frac{e^2 n_{\rm 1D}}{2\pi \epsilon_0} \log(\nu) n_{\rm 2D} .
\end{equation}
We combine the kinetic and potential energies to obtain an energy per electron of 
\begin{equation}
E = \frac{ \pi^2 n_{\rm 1D}^2}{6m} - \frac{e^2 n_{\rm 1D}}{2\pi \epsilon_0}\log(\nu ).
\end{equation}
We consider fixed areal charge density, implying 
\begin{equation}
n_{\rm 1D} = \nu b n_{\rm 2D}.
\end{equation}
This means that holding the total carrier density fixed while increasing the stripe separation requires increasing the charge per stripe. 
We find as a function of $\nu$ 
\begin{equation}
E = \frac{ \pi^2 b^2 n_{\rm 2D}^2 }{6m} \nu^2 - \frac{e^2 b n_{\rm 2D} }{2\pi \epsilon_0}\nu \log(\nu ) =  \frac{ \pi^2 b^2 n_{\rm 2D}^2 }{6m} \left[ \nu^2 -  g \nu \log\nu \right]
\end{equation}
where 
\begin{equation}
g =  \frac{3 m e^2  }{\epsilon_0  \pi^3 b n_{\rm 2D}}
\end{equation}
is the tuning parameter, similar to a Wigner crystal.
We find the ground state energy is minimized for $ 2\nu - g -g\log \nu = 0$.
The solution to this equation exhibits a transition from $\nu = 1$ to $\nu = 2$ for roughly $g \sim 2$. 
At this point, the increased kinetic energy due to the increased Fermi level in the conducting chains is offset by the reduced Coulomb repulsion as the chains undergo a phase separation into stripes.
Increasing $g$ further we see a sequence of further separations into increasingly sparse chains.
The condition of increasing $g$ is affected by reducing the equilibrium carrier density, indicating this is essentially a Wigner crystallization, as argued in~\cite{Spivak.2004,Emery.2000,Schulz.1993}.

We comment that in this case, while the interchain direction remains frozen into a striped crystal, the intrachain dynamics remains coherent in this picture, being essentially regions of electron-rich liquids separated by empty insulating regions.
Therefore transport along the chains is still coherent, barring effects due to disorder which is likely to dramatically reduce this transport. 
It is possible that for sufficiently strong interactions this will also exhibit an instability, e.g. of Peierls type into a checkerboard type insulator, but this probably requires substantial doping in order to obtain the finite density of states. 

Clearly, this problem is very interesting and warrants more attention.
In particular, it is important to develop a better model of the correlations and also account for the kinetic energy in the dispersive direction.
More generally, we see that the "coupled" chain model suggested in~\cite{Wang.2021} is extremely illuminating and if one can dope into this regime, it is likely that low-dimensional strongly-correlated physics will emerge. 
\newpage

%
%
%
%
\centering
\textbf{REFERENCES}
\bibliographystyle{apsrev}
\bibliography{full}